\documentclass[preprint,superscriptaddress,floatfix]{revtex4-2}
\usepackage{graphicx}
\usepackage{footmisc}
\usepackage{subfig,color}
\usepackage{natbib,url}
\usepackage{pifont}
\usepackage{lineno}
\usepackage{amsmath}
\usepackage{float}
\usepackage{placeins}
\usepackage{changepage}
\usepackage{xcolor}

\usepackage[counterclockwise, figuresright]{rotating}

\usepackage{adjustbox}
\usepackage[normalem]{ulem}
\def\be{\begin{equation}}
\def\ee{\end{equation}}

\def\begineqn{\begin{equation*}}
\def\endeqn{\end{equation*}}

\def\beginar{\begin{eqnarray}}
\def\endar{\end{eqnarray}}
\def\beginarn{\begin{eqnarray*}}
\def\endarn{\end{eqnarray*}}

\def\lb{\left ( }
\def\rb{\right ) }
\def\lsq{ [\ }
\def\rsq{ ]\ }

\def\Rat{\widetilde{Ra}}
\def\Ret{\widetilde{Re}}

\def\Nu{Nu}

\def\ist{\ell_{I}^{\vartheta}}
\def\ub{\mathbf{u}}

\def\mth{\overline{\Theta}}

\def\dsx{{\partial_x}}
\def\dsy{{\partial_y}}

\def\dst{{\partial_t}}

\def\dz{{\partial_Z}}

\def\Ret{\widetilde{Re}}

\begin{document}

\title{Small scale quasi-geostrophic convective turbulence at large Rayleigh number}

\author{Tobias G. Oliver}
\affiliation{Department of Physics, University of Colorado, Boulder, CO  80309, USA}
\author{Adrienne S. Jacobi}
\affiliation{Department of Physics, University of Colorado, Boulder, CO  80309, USA}
\author{Keith Julien}
\affiliation{Department of Applied Mathematics, University of Colorado, Boulder, CO  80309, USA}
\author{Michael A. Calkins}
\affiliation{Department of Physics, University of Colorado, Boulder, CO  80309, USA}

\begin{abstract}

	A numerical investigation of an asymptotically reduced model for quasi-geostrophic Rayleigh-B\'enard convection is conducted in which the depth-averaged flows are numerically suppressed by modifying the governing equations.
At the largest accessible values of the Rayleigh number $Ra$, the Reynolds number and Nusselt number show evidence of approaching the diffusion-free scalings of $Re \sim Ra E/Pr$ and $Nu \sim Pr^{-1/2} Ra^{3/2} E^2$, respectively, where $E$ is the Ekman number and $Pr$ is the Prandtl number.
For large $Ra$, the presence of depth-invariant flows, such as large-scale vortices, yield heat and momentum transport scalings that exceed those of the diffusion-free scaling laws.
The Taylor microscale does not vary significantly with increasing $Ra$, whereas the integral length scale grows weakly. 
The computed length scales remain $O(1)$ with respect to the linearly unstable critical wavenumber; we therefore conclude that these scales remain viscously controlled. 
We do not find a point-wise Coriolis-Inertia-Archimedean (CIA) force balance in the turbulent regime; interior dynamics are instead dominated by horizontal advection (inertia), vortex stretching (Coriolis) and the vertical pressure gradient. 
A secondary, sub-dominant balance between the Archimedean buoyancy force and the viscous force occurs in the interior and the ratio of the rms of these two forces is found to approach unity with increasing $Ra$. This secondary balance is attributed to the turbulent fluid interior acting as the dominant control on the heat transport. These findings indicate that a pointwise CIA balance does not occur in the high Rayleigh number regime of quasi-geostrophic convection in the plane layer geometry.
Instead, simulations are characterized by what may be termed a \textsl{non-local} CIA balance in which the buoyancy force is dominant within the thermal boundary layers and is spatially separated from the interior Coriolis and inertial forces.

\end{abstract}

\maketitle

%\keywords{Multiscale models, quasi-geostrophy, convection}%Use showkeys class option if keyword

\section{Introduction}
\label{S:intro}

%\textcolor{red}{Things to put in intro: motivation for study -- describe importance of rotating convection, difficulties associated with studying, asymptotic regime of rotating convection, LSVs and their influence, asymptotic regime of turbulent rotating convection, motivate the reason for removing the LSV}

%Primary goals: understand scaling behavior of heat and momentum transport in high Rayleigh number limit; understand variation of length scales with increasing level of turbulence.
%
%P: motivation
%P: define system
%P: properties of rotating convective turbulence: length scales
%P: inverse cascade
%P: scaling of heat and momentum transport
%P: summary of paper

Convection plays an important role in the dynamics of many planets and stars, where it serves as the power source for sustaining the magnetic fields of the planets \citep{sS10,cJ11b,pR13,jmA15} and the Sun \citep{pC14}. Convection may also be a possible driving mechanism for the observed large scale zonal winds  \citep[e.g.][]{mH16} and vortices \citep[e.g.][]{lS22} on the giant planets. The flows in these natural systems are strongly forced and turbulent and can be constrained by the Coriolis force. 
%Indeed, the Coriolis force is thought to be important for the formation of both global scale magnetic field \citep[e.g.][]{eP55,mS66a,hM72,sC72} and zonal winds \citep[e.g.][]{fB02}. 
Studying rotationally constrained convective turbulence is therefore important for improving understanding of such systems.
However, experimental \citep{rE14,jC15,mM21,tV21} and numerical \citep{tG16,aG21,mY22} investigations have difficulty accessing this parameter regime due to the extreme scale separation that characterizes the dynamics.
%but is intrinsically difficult due to the wide range of spatiotemporal scales that characterize such flows. Laboratory experiments and direct numerical simulations are all used toward this end, but extrapolation of such results to natural systems is questionable due to the vast separation in parameter values between model and nature. 
Asymptotic models play an important role in this regard since they allow for significant computational savings by eliminating physically unimportant dynamics while retaining the dominant force balance that is thought to be representative of natural systems.
In particular, the asymptotic model for rapidly rotating convection in a planar geometry has been used to advance our understanding of this system \citep{kJ98a}, and shows excellent agreement with the results of DNS where an overlap of the parameter space is possible \citep{sS14,mP16}. Nevertheless, the behavior of the system in the dual limit of strong buoyancy forcing \textit{and} rapid rotation is still not completely understood \citep{sM21}.
In the present work we use numerical simulations of this asymptotic model for investigating the scaling behavior of rotating convective turbulence at previously inaccessible parameter values.

%the strongly forced, geostrophic turbulence regime in which the depth-averaged large scale flows are suppressed numerically. Suppression of these large scale flows provides computational savings that allow for a broader range of parameter values to be simulated. 

%we However, the identification of asymptotic scaling behavior in model output parameters is then paramount for extrapolating model results to natural systems.
%
%
%Laboratory experiments, direct numerical simulations, and asymptotic models are all used toward this end, but extrapolation of such results to natural systems is questionable due to the vast separation in parameter values between model and nature. 
%
%Identifying asymptotic scaling behavior in model output parameters is then paramount for extrapolating model results to natural systems \citep[e.g.][]{jmA07}. 

Rayleigh-B\'enard convection, consisting of a fluid layer of depth $H$ confined between plane parallel boundaries is a canonical system used for studying buoyancy-driven flows. The two boundaries have temperature difference $\Delta T$ and a constant gravitational field of magnitude $g$ points perpendicular to the boundaries. The Rayleigh number 
\be
Ra = \frac{g \alpha \Delta T H^3}{\nu \kappa},
\ee
provides a non-dimensional measure of the buoyancy force. Here, $\alpha$ is the thermal expansion coefficient, $\nu$ is the kinematic viscosity and $\kappa$ is the thermal diffusivity. In most systems of interest, the flow is strongly driven such that \textcolor{black}{$Ra \gg \textcolor{black}{Ra_{c},}$ where $Ra_c$ is the critical Rayleigh number for the onset of convection.} 
In the presence of a background rotation with angular frequency $\Omega$, the resulting convective dynamics are considered rotationally constrained, or quasi-geostrophic (QG), when viscous forces and inertia are both small relative to the Coriolis force. In non-dimensional terms, quasi-geostrophy is characterized by small Rossby and Ekman numbers, respectively defined by,
\be
Ro = \frac{U}{2 \Omega H }, \qquad E = \frac{\nu}{2 \Omega H^2},
\ee
where $U$ is a characteristic flow speed. In addition, the Reynolds number is given by $Re = U H/ \nu = Ro/E$; QG turbulence is characterized by $Re \gg 1$, and therefore $E \ll Ro \ll 1$. As an example, for the Earth's outer core estimates suggest $Ro = O(10^{-7})$, $E = O(10^{-15})$ and $Re = O(10^8)$ \citep[e.g.][]{pR13}. 

%Understanding how heat and momentum transport, as characterized by the Nusselt number $Nu$ and Reynolds number, respectively, behave in the high-Rayleigh number regime of QG convection is vital for relating model output to natural systems \textcolor{red}{This is kind of an awkward sentence. How about:}

An inverse kinetic energy cascade is known to occur in QG convection \citep{kJ12,aR14,bF14,cG14,sM21}. In the periodic plane layer geometry, the inverse cascade gives rise to a depth-invariant large scale vortex (LSV) that tends to equilibrate with lateral dimensions comparable to the horizontal size of the simulation domain. The LSV can then dynamically influence the underlying small scale convection. Previous studies have shown that the relative amount of kinetic energy contained in the LSV and the small scale convection depends non-monotonically on the forcing amplitude \citep{cG14,sM21}. In addition, the characteristic speed of the LSV depends linearly on the aspect ratio of the simulation domain \citep{sM21,jN22}. Although such coupling between the large and small scale dynamics is directly relevant to natural systems, it nevertheless complicates efforts to understand the strongly forced asymptotic regime of the small scale convection.

Momentum and heat transport are \textcolor{black}{characterized} by $Re$ and the Nusselt number, $Nu$, respectively. Understanding the scaling of these two quantities in the high Rayleigh number regime of QG convection is vital for relating model output to natural systems.
QG convection dynamics are conveniently specified by the asymptotic combination 
\be
\Rat = Ra E^{4/3},
\ee
which we refer to as the reduced Rayleigh number. With this rescaling, the onset of convection occurs at $\Rat \approx 8.7$, and the onset of turbulence occurs when $\Rat \gtrsim 40$  \citep{kJ12}. Previous work has found heat transport data that is consistent with the diffusion-free scaling, $Nu \sim \Rat^{3/2}Pr^{-1/2}$, in the turbulent regime \citep{kJ12b,tG16}, where $Pr=\nu/\kappa$ is the Prandtl number. 
However, recent investigations that allow for the amplitude of the LSV to fully saturate have found a stronger scaling for $Nu$ \citep{sM21}, which might suggest that although the small scale convection approaches a $Nu \sim \Rat^{3/2}Pr^{-1/2}$, the interaction with the LSV provides an additional enhancement of heat transport.
%, possibly by generating a coherent box scale convective mode. 
A similar diffusion-free scaling for the flow speeds is given by $Re \sim Ra E/Pr$ implying a rotational free-fall velocity  $U_{r\!f\!f} = g \alpha \Delta T/2\Omega$.  Recent studies of convection in a spherical geometry support this behavior \citep{cG19}. However, similar to their heat transport findings, Ref.~\cite{sM21} find that the Reynolds number scales more strongly than the diffusion-free scaling. 

%One of the goals of the present investigation is to isolate the scaling behavior of the small scale convection by suppressing the growth of the LSV. 

%The Reynolds number is given by $Re = U H/ \nu = Ro/E$. The non-dimensional regime in which $E \ll Ro \ll 1$ \textit{and} $Re \gg 1$ can be used to define the parameter space of geostrophic convective turbulence \citep[e.g.][]{kJ12}, and natural systems are thought to lie within this regime \citep[][]{jmA15}. Thus, studies of QG convection often focus on identifying asymptotic scaling behavior in various physical quantities such that model results can be extrapolated to the regime of natural systems. 

%Some of these quantities include heat and momentum transport.

%Efforts to understand this regime are particularly focused on the identification of asymptotic behavior since such  

%The quasi-geostrophic model developed by \cite{kJ98a} is constructed to reside within this regime.

A defining property of turbulence is the presence 
of a broad range of length scales within the flow field. In spectral space this range of scales translates into a broadband kinetic energy spectrum. 
The spectrum of non-rotating turbulence broadens via a forward cascade in which there is a net transfer of energy from low wavenumbers to high wavenumbers. A consequence of this transfer of energy is that the length scale at which viscous dissipation is dominant becomes ever smaller as $Re$ increases \citep[e.g.][]{sP00}. 
%This phenomenology is generally consistent with studies of non-rotating convection \citep[e.g.][]{mY21}.

The process by which spectral broadening occurs in QG convection is less clear. 
In the limit $E \rightarrow 0$, linear theory shows that the onset of convection occurs on a length scale of size $\ell = O\lb E ^{1/3}\rb$.
%Linear theory shows that at the onset of convection the flow is forced on a length scale of size $O(E^{1/3})$ in the limit $E \rightarrow 0$. 
This length scale arises because the viscous force facilitates convection by simultaneously perturbing the geostrophic force balance and relaxing the Taylor-Proudman constraint. 
Understanding which length scales emerge in the strongly nonlinear regime is important for characterizing QG convective turbulence. 
Previous studies \textcolor{black}{in spherical geometry} suggest that the length scale varies with the Rossby number as $\ell \sim Ro^{1/2}$ \citep[e.g.][]{cG19}, which is thought to arise from the so-called Coriolis-Inertia-Archimedean (CIA) balance \citep{cJ15}. In terms of the reduced Rayleigh number this scaling is equivalent to $\ell \sim E^{1/3} \sqrt{\Rat/Pr}$ \citep[e.g.][]{jmA20}. 
Recent experimental work using water as the working fluid has found a slightly weaker scaling than $\ell \sim Ro^{1/2}$ \citep{mM21}; this same investigation, and a numerical study of convection-driven dynamos \citep{mY22}, finds that the correlation length scale of the vorticity is approximately constant with increasing $\Rat$. It is presently unknown whether this behavior persists in the limit of large $\Rat$.

In the present investigation we report on the results of numerical simulations of the QG model of rotating convection in which the depth-invariant flows are suppressed. This suppression is done to isolate the asymptotic behavior of the small-scale convection in the absence of the LSV, and to simulate the largest accessible values of $\Rat$ in an attempt to identify asymptotic trends in the scaling behavior of various flow quantities. In Sec.~\ref{S:methods} we provide an overview of the QG model and numerical methods. Results and conclusions are given in  Sec.~\ref{sec:results} and Sec.~\ref{S:conclusions}, respectively.

\section{Methods}
\label{S:methods}

\subsection{Governing equations}

%We consider a rotating layer of Boussinesq fluid with depth $H$, bounded by plane parallel boundaries. The rotation rate of the system is $\Omega$, the gravitational acceleration is $g$, and the constant temperature difference between the boundaries is denoted by $\Delta T$. The fluid has kinematic viscosity $\nu$, thermal diffusivity $\kappa$ and thermal expansion coefficient $\alpha$. 

In the present investigation we employ \textcolor{black}{a modified version of} the asymptotically reduced form of the governing equations given by Ref.~\citep[][]{mS06}. When non-dimensionalized using the small-scale viscous diffusion speed $\nu/\ell^*_\nu$, where $\ell^*_\nu = H E^{1/3}$, \textcolor{black}{and temperature scale $\Delta T$,} these equations take the form
	\be
\dst \zeta + J\lsq \psi,\zeta \rsq - \textcolor{black}{\gamma \left< J\lsq \psi,\zeta \rsq \right>} - \dz w = \nabla_\perp^2 \zeta,
		\label{eqn:barotropicvort}
	\ee
	\be
		\dst w + J\lsq \psi , w \rsq + \dz \psi = \frac{\Rat}{Pr}\vartheta + \nabla_\perp^2 w ,
		\label{eqn:barotropicmom}
	\ee
	\be
		\dst \vartheta + J\lsq \psi , \vartheta \rsq + w \dz \mth = \frac{1}{Pr} \nabla_\perp^2 \vartheta,
		\label{eqn:barotropictemp}
	\ee
	\be
		\dz \lb \overline{w \vartheta} \rb = \frac{1}{Pr} \partial_Z^2 \mth,
		\label{eqn:barotropicdff}
	\ee
%	\begin{equation}
%		\nabla_\perp \cdot\mathbf{u}_\perp^{ag} + \dz w = 0
%		\label{eqn:incompressible}
%	\end{equation}
	where $t$ is time, the Cartesian coordinate system is denoted by $(x,y,Z)$, the Jacobian operator is defined by $J\lsq \psi ,A\rsq = \dsx \psi \dsy A -\dsy \psi \dsx A$ for some scalar field $A$, \textcolor{black}{$\gamma$ is a constant, the angled brackets appearing in equation \eqref{eqn:barotropicvort} denote an average over the depth ($Z$)}, and the horizontal Laplacian operator is denoted by $\nabla_\perp^2 = \partial_x^2 + \partial_y^2$. The vertical components of vorticity and velocity are denoted by $\zeta$ and $w$, respectively, $\psi$ is the geostrophic streamfunction, and $\vartheta$ is the fluctuating temperature.  The vorticity and streamfunction are related via $\zeta = \nabla_\perp^2 \psi$. The mean temperature is \textcolor{black}{denoted by} $\mth$, where the overline represents a horizontal average. \textcolor{black}{We note that $\mth = O(1)$ and $\vartheta=O(E^{1/3})$ in the asymptotic expansion. }

	\textcolor{black}{ 
The constant $\gamma$ is either one or zero. When $\gamma = 0$ the above equations are identical to those used in many previous investigations \citep[e.g.][]{sM21}. For simulations in which the depth-averaged flow is suppressed we set $\gamma = 1$; in this case a depth-average of equation \eqref{eqn:barotropicvort} yields the diffusion equation
\be
\dst \left<\zeta\right>= \nabla_{\perp}^{2}\left<\zeta\right> ,
\ee
so that the depth-averaged vorticity (streamfunction) trivially decays to zero as $t \rightarrow \infty$. In practice, simulations in which the depth-averaged flow is suppressed were initialized from states in which  $\left<\psi\right> = 0$.}

\textcolor{black}{
Simulations are performed for the range $10 \le \Rat \le 280$ with $Pr=1$.
Details of the numerical simulations are provided in Table \ref{tab:run}. The equations are solved using a de-aliased pseudo-spectral method in which the flow variables are expanded as Chebyshev polynomials in the vertical dimension and Fourier series in the horizontal dimensions. A third-order accurate implicit-explicit Runge-Kutta scheme is used to advance the equations in time. Further details on the code can be found in Ref.~\cite{pM16}.
}
%\begin{figure}
%	\begin{center}
%		\includegraphics[width=0.7\textwidth]{ke}
%	\end{center}
%	\caption{Kinetic energy along the three directions of flow for a representative case of $\Rat = 160$. Data from the simulation in which the depth-averaged flow is present is denoted by the dashed lines, while the simulation in which in the depth-averaged flow is suppressed is shown by the solid lines. Note that the horizontal kinetic energy components are significantly reduced when the depth-averaged flow is suppressed, though the vertical component of the kinetic energy is comparable for the two cases.}
%	\label{f:ke}
%\end{figure}

\begin{figure}
	\begin{center}
		\includegraphics[width=0.6\textwidth]{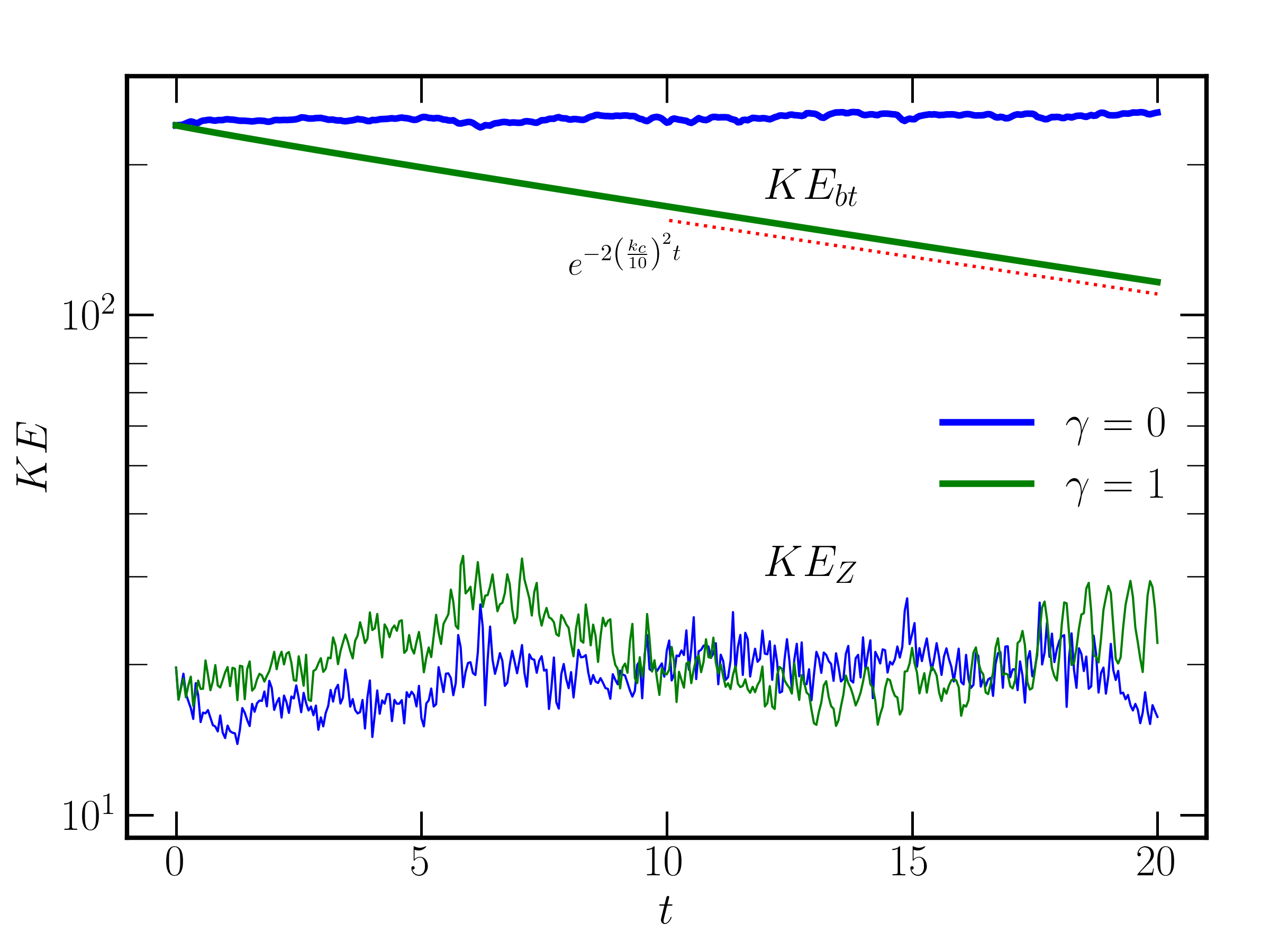}
	\end{center}
	\caption{Temporal evolution of the depth-averaged (barotropic, thick lines) and vertical (thin lines) kinetic energy densities for two simulations with $\Rat = 100$, illustrating the effect of suppressing the depth-averaged component of the flow. The green and blue data correspond to the model where the depth-averaged flow is suppressed ($\gamma=1$) and maintained ($\gamma=0$), respectively. Both models were initialized from the same initial state. The dotted red line is the expected, long-time scaling for the suppressed model where $k_c = 1.3048$ is the critical wavenumber.}
	\label{figure:ke}
\end{figure}

\textcolor{black}{As a demonstration of the approach for suppressing the depth-averaged flow, figure \ref{figure:ke} shows the temporal evolution of the volume averaged kinetic energy density for two sample simulations with $\Rat=100$. The simulation in which the depth-averaged flow is suppressed (maintained) is denoted by $\gamma=1$ ($\gamma=0$). Both simulations were initialized with identical initial conditions. We compute the depth-averaged (barotropic, $KE_{bt}$) and vertical ($KE_Z$) kinetic energy densities, respectively defined as
\be
KE_{bt} = \frac{1}{2 A} \int \lsq \langle u \rangle^2 + \langle v \rangle^2 \rsq dA , \qquad
KE_{Z} = \frac{1}{2 V} \int w^2 dV ,
\ee
where the horizontal velocity components are denoted by $(u,v) = \lb - \dsy \psi, \dsx \psi \rb$, $A$ is the area of a horizontal cross section and $V$ is the volume.
We observe a clear exponential decay of $KE_{bt}$ when $\gamma=1$; the predicted decay rate for the dipolar vortex is shown for comparison and excellent agreement is observed.}

\subsection{Diagnostic quantities}
Heat and momentum transport are quantified with the Nusselt number $Nu$ and the asymptotically reduced Reynolds number $\Ret$, respectively. The Nusselt number is defined by 
\begin{equation}
	Nu = 1 + Pr\left<w \vartheta\right>,
	\label{E:Nu}
\end{equation}
where the angled brackets denote a volume and time average. The asymptotically rescaled Reynolds number is defined based on the vertical component of the velocity field such that
\begin{equation}
	\Ret = E^{1/3}Re = \sqrt{\left<w^{2}\right>}.
	\label{E:Re}
\end{equation}

For future reference, the relationship
\be
\frac{\Rat}{Pr^2} \lb Nu - 1 \rb = \varepsilon_u,
\quad
Nu = \varepsilon_\theta
\label{eq:edisspn} 
\ee
\textcolor{black}{can be derived from the governing equations under the condition $\gamma = 0$ or if the initial condition is such that $\left<\psi\right> = 0$ \citep{kJ12},} where the viscous and thermal dissipation rates are defined by, respectively,
\be
\label{eq;powerlaw}
\varepsilon_u = \langle | \nabla_\perp w|^2\rangle + \langle \zeta^2\rangle, \quad
\varepsilon_\theta =  \langle | \nabla_\perp \theta|^2\rangle + \langle \lb \partial_Z\overline{\Theta}\rb^2\rangle.
\ee

The integral scale, Taylor microscale, Kolmogorov scale, and temperature integral scale are calculated for each simulation and defined by, respectively,
\begin{equation}	
\ell_I=\frac{\int k^{-1} \widehat{E}_{u}(k) dk}{\int \widehat{E}_{u}(k) dk}, \quad 
\ell_{T} = \lb \frac{  \int \widehat{E}_{u}(k) dk}{\int k^2  \widehat{E}_{u}(k) dk} \rb^{1/2} , \quad
\ell_K = \varepsilon_u^{-1/4}, \quad
\ell_I^{\vartheta}=\frac{\int k^{-1} \widehat{E}_{\vartheta}(k) dk}{\int \widehat{E}_{\vartheta}(k) dk} 
\label{eqn:scales}
\end{equation}
where the time- and depth-averaged kinetic energy (temperature variance) spectrum is denoted by $\widehat{E}_u(k)$ ($\widehat{E}_{\vartheta}\lb k \rb$), and the horizontal wavenumber vector is $\mathbf{k} = (k_x, k_y)$ with modulus $k = \sqrt{k_x^2 + k_y^2}$. 

The Kolmogorov scale is often interpreted as the scale at which viscosity dominates and the turbulent kinetic energy is dissipated into heat \textcolor{black}{\citep{gV06}}. 
The integral scale and Taylor microscale are respectively interpreted as measures of the correlation length of turbulent motions and the intermediate length scale at which fluid viscosity significantly affects the dynamics of turbulent motions \textcolor{black}{\citep{sP00}}. 
With these interpretations in mind, the relative ordering $\ell_I > \ell_T> \ell_K$ holds.
Furthermore, for rotating convection, viscosity is a required ingredient in destabilizing a rotating fluid subject to an adverse temperature gradient \citep{sC61}. 
This implies that convective motions are inherently influenced by viscosity and, as a consequence, the Taylor microscale (assuming the physical interpetation holds) is tied to the linear instability scale \textcolor{black}{such that we anticipate} $\ell_{T} = O(1)$. The integral scale for the temperature is interpreted as the scale at which the fluid is forced by buoyancy.

%\textcolor{red}{The Kolmogorov scale defines the scale at which viscosity dominates and the turbulent kinetic energy is dissipated into heat. The integral scale and Taylor microscale  respectively measure the correlation length of turbulence and the intermediate length scale at which fluid viscosity significantly affects the dynamics of turbulent motions. Hence,  by definition the relative ordering $\ell_I > \ell_T> \ell_K$ holds. Furthermore, for rotating convection viscosity is a required ingredient in destabilizing a rotating fluid subject to an adverse temperature gradient \citep{sC61}. This implies that  convective motions are inherently influenced by viscosity, as a consequence the Taylor microscale is tied to the linear instability scale,  i.e., $\ell_T\sim1$.  }
%
%\textcolor{red}{TO- I am thinking that maybe we should say that $\ell_{I}$, $\ell _{T}$ \textit{are often interpreted as the scales at which...}, and that this interpretation would suggest that $\ell_{T}\sim 1.$}
%
%\textcolor{red}{KJ- I would also define and plot $\ell^\theta_I$ using $\hat{E}_{\theta,rms}$. Why? 1) buoyancy is the driver of all motions so this should also be sensitive to, or indeed, drive changes in the integral correlation length. 2) This is something experimentalist can measure accurately, we should lead the charge by proposing this,  ...}.

\setlength{\tabcolsep}{2mm}
\renewcommand{\arraystretch}{0.5}
\begin{table}
		%\begin{footnotesize}
	\begin{tabular}{c c c c c c c c c}
	%\hline 
		$\Rat$& $N_x\times N_y\times N_Z$ & $\Delta t$ & $n$ & $\Ret$ &  $Nu$ &$\ell_{I}$ & $\ell_{T}$ &$\ell_{K}$ \\
		\hline
		$10$ & $384\times384\times96$ & $5.0\times 10^{-4}$ & $10$ & $0.592$ & $1.259$ & $1.081$ & $1.076$&$0.899$\\
$12$ & $384\times384\times96$ & $5.0\times 10^{-4}$ & $10$ & $0.921$ & $1.602$ & $1.165$ & $1.146$&$0.687$\\
$15$ & $384\times384\times96$ & $5.0\times 10^{-4}$ & $10$ & $1.405$ & $2.132$ & $1.266$ & $1.203$&$0.554$\\
$20$ & $384\times384\times96$ & $5.0\times 10^{-4}$ & $10$ & $2.531$ & $3.349$ & $1.35$ & $1.219$&$0.437$\\
$25$ & $384\times384\times96$ & $5.0\times 10^{-4}$ & $10$ & $4.247$ & $5.269$ & $1.41$ & $1.191$&$0.364$\\
$30$ & $384\times384\times96$ & $2.5\times 10^{-4}$ & $10$ & $6.008$ & $7.349$ & $1.495$ & $1.185$&$0.319$\\
$40$ & $384\times384\times96$ & $2.5\times 10^{-4}$ & $10$ & $9.556$ & $11.61$ & $1.655$ & $1.159$&$0.266$\\
$50$ & $432\times432\times144$ & $2.5\times 10^{-4}$ & $10$ & $13.09$ & $16.3$ & $1.793$ & $1.119$&$0.233$\\
$60$ & $576\times576\times144$ & $2.5\times 10^{-4}$ & $10$ & $15.58$ & $19.81$ & $1.75$ & $1.035$&$0.215$\\
$70$ & $576\times576\times180$ & $2.0\times 10^{-4}$ & $10$ & $17.91$ & $22.88$ & $1.676$ & $0.958$&$0.201$\\
$80$ & $576\times576\times192$ & $2.0\times 10^{-4}$ & $10$ & $20.38$ & $26.25$ & $1.723$ & $0.937$&$0.187$\\
$100$ & $648\times648\times288$ & $2.0\times 10^{-4}$ & $10$ & $26.99$ & $35.66$ & $1.921$ & $0.938$&$0.164$\\
$120$ & $768\times768\times324$ & $1.0\times 10^{-4}$ & $10$ & $34.66$ & $46.90$ & $2.199$ & $0.955$&$0.146$\\
$140$ & $720\times720\times324$ & $1.0\times 10^{-4}$ & $10$ & $43.22$ & $60.81$ & $2.407$ & $0.963$&$0.131$\\
$160$ & $768\times768\times360$ & $1.0\times 10^{-4}$ & $10$ & $54.05$ & $77.65$ & $2.725$ & $0.995$&$0.119$\\
$180$ & $810\times810\times384$ & $5.0\times 10^{-5}$ & $10$ & $63.12$ & $95.73$ & $2.829$ & $0.991$&$0.109$\\
$200$ & $960\times960\times450$ & $5.0\times 10^{-5}$ & $10$ & $69.91$ & $110.8$ & $2.796$ & $0.965$&$0.102$\\
$220$ & $960\times960\times480$ & $2.5\times 10^{-5}$ & $10$ & $82.18$ & $135.0$ & $3.0$ & $0.983$&$0.095$\\
$240$ & $1125\times1125\times540$ & $2.5\times 10^{-5}$ & $10$ & $89.00$ & $150.8$ & $2.968$ & $0.962$&$0.090$\\
$260$ & $1200\times1200\times600$ & $2.0\times 10^{-5}$ & $10$ & $95.81$ & $165.7$ & $2.988$ & $0.948$&$0.086$\\
$280$ & $1200\times1200\times675$ & $2.0\times 10^{-5}$ & $10$ & $104.6$ & $184.7$ & $3.028$ & $0.941$&$0.083$\\
$100$ & $324\times324\times288$ & $2.0\times 10^{-4}$ & $5$ & $26.37$ & $34.38$ & $1.662$ & $0.926$&$0.165$\\
$100$ & $480\times480\times288$ & $2.0\times 10^{-4}$ & $7$ & $26.66$ & $35.21$ & $1.879$ & $0.937$&$0.164$\\
$100$ & $768\times768\times288$ & $2.0\times 10^{-4}$ & $12$ & $26.78$ & $35.52$ & $2.06$ & $0.948$&$0.164$\\
$100$ & $972\times972\times288$ & $2.0\times 10^{-4}$ & $15$ & $26.93$ & $36.92$ & $3.908$ & $1.142$&$0.164$\\
$100$ & $1296\times1296\times288$ & $2.0\times 10^{-4}$ & $20$ & $27.11$ & $35.92$ & $1.943$ & $0.937$&$0.163$\\
$40$ & $192\times192\times96$ & $2.5\times 10^{-4}$ & $5$ & $9.364$ & $10.96$ & $1.640$ & $1.169$&$0.271$\\
$40$ & $288\times288\times96$ & $2.5\times 10^{-4}$ & $7$ & $9.25$ & $11.12$ & $1.650$ & $1.166$&$0.269$\\
$40$ & $480\times480\times96$ & $2.5\times 10^{-4}$ & $12$ & $9.561$ & $11.46$ & $1.686$ & $1.172$&$0.267$\\
$40$ & $576\times576\times96$ & $2.5\times 10^{-4}$ & $15$ & $9.447$ & $11.41$ & $1.666$ & $1.168$&$0.267$\\
$40$ & $768\times768\times96$ & $2.5\times 10^{-4}$ & $20$ & $9.564$ & $11.47$ & $1.729$ & $1.178$&$0.266$\\

	\end{tabular}
	\caption{Details of the numerical simulations in which depth-invariant flows are suppressed. The number of physical space grid points in each respective direction is specified by $N_x\times N_y\times N_Z$, the time step size is denoted by $\Delta t$, and the horizontal domain size, as specified by the integer number of critical horizontal wavelengths, is denoted by $n$. The asymptotically rescaled Reynolds number is $\Ret$, $Nu$ is the Nusselt number, and the integral length scale, Taylor microscale and Kolmogorov scale are denoted by $\ell_I$, $\ell_T$ and $\ell_K$, respectively. All length scales are normalized by the critical wavelength $\lambda_{c} \approx 4.815$. }
	\label{tab:run}
\end{table}
\begin{table}
		%\begin{footnotesize}
	\begin{tabular}{c c c c c c c c c}
	%\hline 
 %& & & LSV& & & & & \\
		$\Rat$& $N_x\times N_y\times N_Z$ & $\Delta t$ & $n$ & $\Ret$ &  $Nu$ &$\ell_{I}$ & $\ell_{T}$ &$\ell_{K}$ \\
		\hline
$20$ & $384\times384\times96$ & $5\times 10^{-4}$ & $10$ & $3.541$ & $4.01$ & $2.048$ & $1.189$&$0.414$\\
$30^{*}$ & $128\times128\times64$ & $5\times 10^{-4}$ & $10$ & $7.219$ & $7.960$ & $--$ & $--$&$--$\\
$40$ & $384\times384\times96$ & $5\times 10^{-4}$ & $10$ & $10.59$ & $11.79$ & $1.705$ & $1.023$&$0.272$\\
$60^{*}$ & $256\times256\times96$ & $1\times 10^{-4}$ & $10$ & $16.82$ & $19.96$ & $--$ & $--$&$--$\\
$80$ & $576\times576\times192$ & $1\times 10^{-4}$ & $10$ & $24.68$ & $30.92$ & $2.291$ & $0.928$&$0.186$\\
$120$ & $648\times648\times324$ & $1\times 10^{-4}$ & $10$ & $41.4$ & $58.2$ & $2.805$ & $0.974$&$0.140$\\
$160$ & $768\times768\times384$ & $5\times 10^{-5}$ & $10$ & $59.40$ & $98.06$ & $2.792$ & $0.919$&$0.118$\\
$200$ & $960\times960\times450$ & $5\times 10^{-5}$ & $10$ & $84.21$ & $146.2$ & $3.032$ & $0.943$&$0.098$\\
		\hline

	\end{tabular}
		%\end{footnotesize}
	\caption{Details of the numerical simulations for cases in which depth-invariant flows are not suppressed. See Table \ref{tab:run} for specifics. Runs denoted with an asterisk ($*$) are from Ref.~\cite{sM21}.}
\end{table}

\section{Results}
\label{sec:results}

\subsection{Influence of simulation domain size}
 
\begin{figure}
	\subfloat[]{\includegraphics[width=0.32\textwidth]{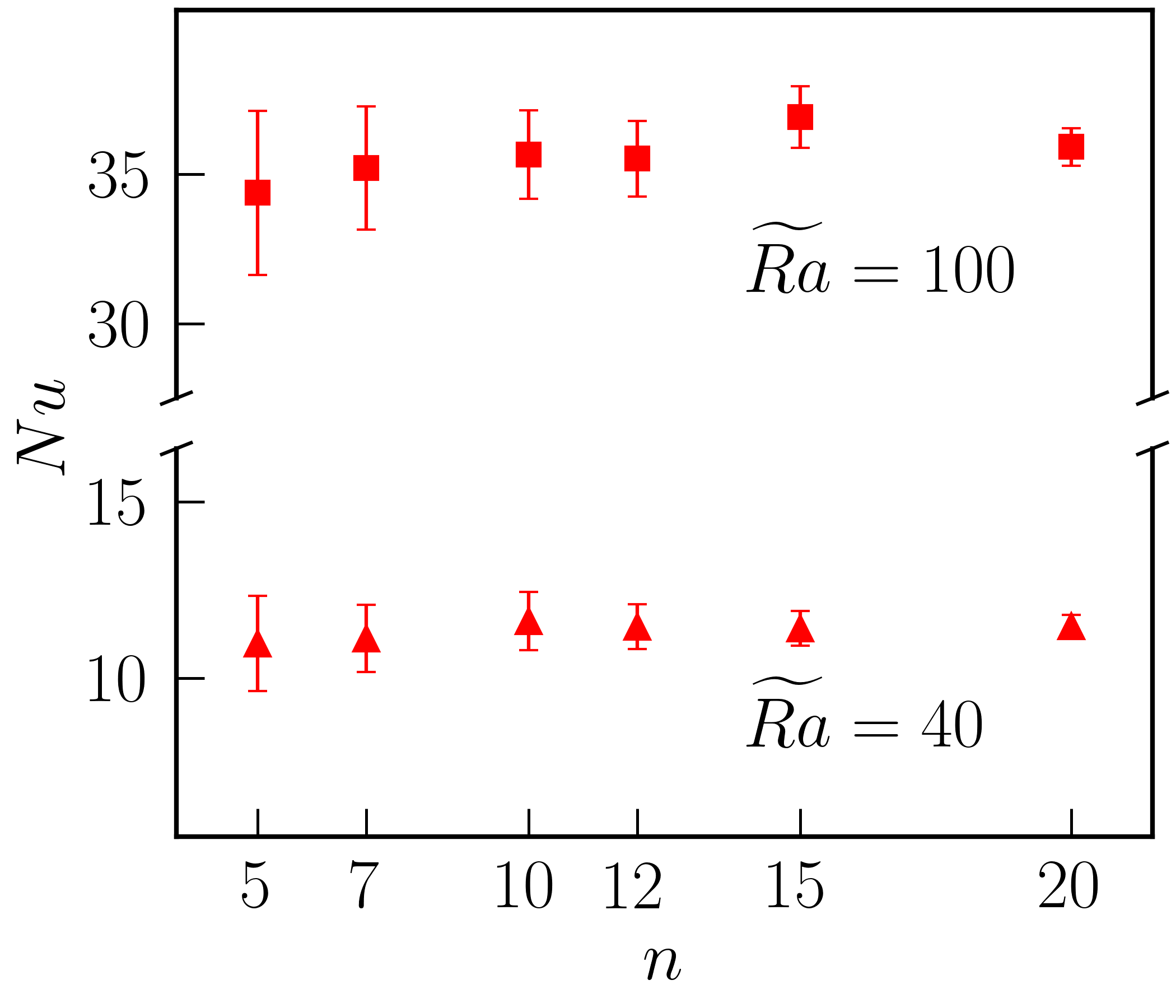}
\label{sub:box_nondim}}\hspace{1mm}
\subfloat[]{\includegraphics[width=0.32\textwidth]{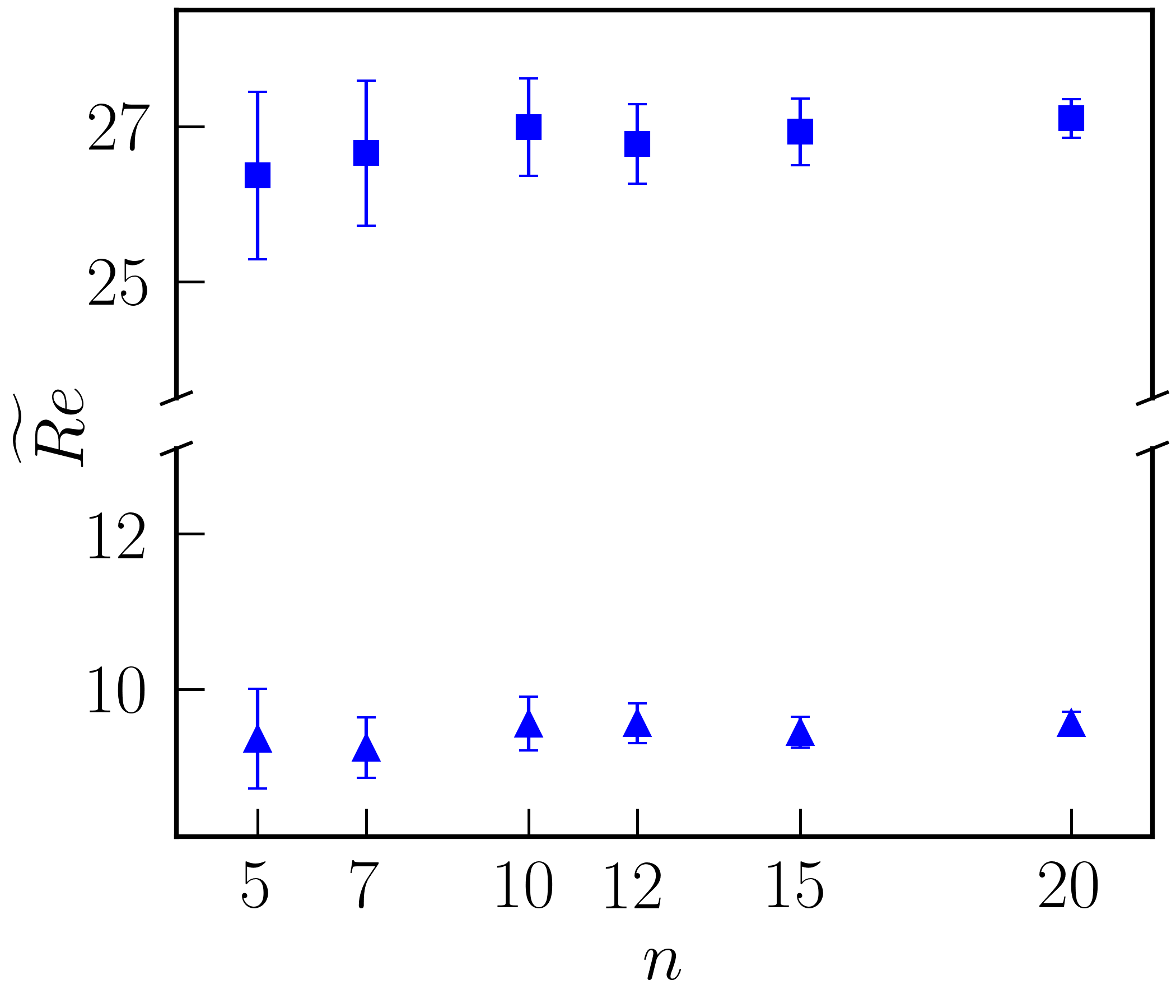}\label{sub:box_reynolds}}\hspace{1mm}
\subfloat[]{\includegraphics[width=0.32\textwidth]{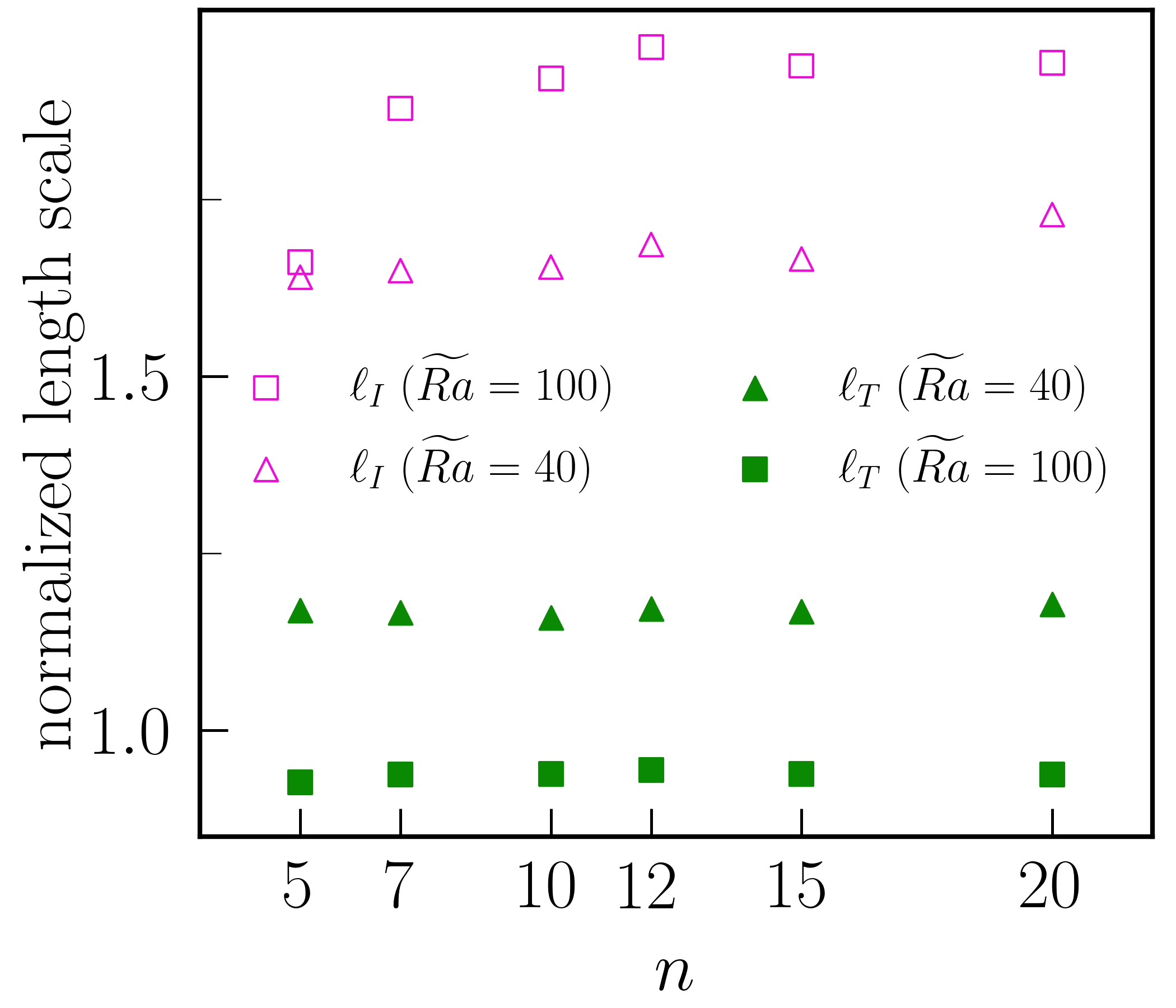}\label{sub:box_ls}}

	\caption{Various flow quantities as a function of horizontal domain size (as characterized by the number of critical horizontal wavelengths, $n$) for $\Rat =100$ (\ding{110}) and $\Rat =40$ (\ding{115}). Error bars denote the standard deviation of globally averaged quantities. 
		\protect\subref{sub:box_nondim} Nusselt number; \protect\subref{sub:box_reynolds} Reynolds number; (c) integral length scale ($\ell_I$) and Taylor microscale ($\ell_T$). 
	}
		%The values of both non-dimensional parameters are robust to changes in the horizontal dimensions of the box. 
	%Variations in the Taylor microscale appear robust to changes in the box size. 
	%For $\Rat = 100$, the integral scale does not asymptote, but remains $O\lb 1 \rb$. 
	%\textcolor{red}{The reasoning for this non-convergent behavior may be due to the coarse resolution in $k$ space at low $k$.}
	%This suggests that the simmulation domain of $10\lambda_{c}$ used in this study is sufficient. 
	\label{figure:box_dim}
\end{figure}

The presence of an inverse cascade complicates our understanding of rotating convection since the flow speeds associated with the LSV are known to grow linearly with the horizontal dimension of the simulation domain size \citep{sM21}. This linear dependence is tied to the fact that the inverse cascade is halted solely by the viscous force acting on the domain scale \citep{jN22}. While we eliminate the LSV (and all depth-invariant motions) in the current investigation, it is nevertheless important to determine the domain size that allows for convergence of key statistical quantities. 
Towards this end, a series of simulations with $\Rat=40$ and $\Rat=100$ were performed for varying domain sizes. Here we scale the horizontal dimension of the simulation domain in integer multiples ($n$) of the critical horizontal wavelength, $\lambda_c \approx 1.3048$.
The effective resolution (i.e.~number of grid points per critical wavelength) was held approximately constant as the domain size was varied.
Figure \ref{figure:box_dim} shows the sensitivity of $\Nu$, $\Ret$, and the length scales $\ell_{I}$ and $\ell_{T}$ as a function of $n$. We find that $Nu$, $\Ret$, and $\ell_T$  show no significant sensitivity to domain size with $n \ge 10$, which is consistent with Ref.~\citep{sM21}. 
\textcolor{black}{The integral scale and the Taylor microscale also show little sensitivity to the box size beyond $n=10$.}
%We find that the integral length scale $\ell_I$ exhibits more sensitivity to the domain size. For $\Rat=40$, it appears that $\ell_I$ converges to a nearly constant value for $n \ge 12$. For $\Rat=100$ it is less clear whether convergence in $\ell_I$ is achieved for the simulated range of $n$. 
The focus of the present investigation is to reach the largest computationally affordable values of $\Rat$; for this reason we choose $n=10$ for all of the data that is presented in later subsections. 
%Moreover, we find that the differences in $\ell_I$ between simulations with $n=10$ and $n > 10$ are less than $10 \%$ over our investigated range of parameters.

%\textcolor{red}{Does the integral length scale for the temperature converge with increasing $n$?}

\subsection{Heat and momentum transport}
\label{S:Nu_Re}

\begin{figure}
	\centering
%\subfloat[]{\includegraphics[width=0.45\textwidth]{Nu_Plot.pdf}}%\hspace{-0.6cm}
%\quad
%\subfloat[]{\includegraphics[width=0.45\textwidth]{Re_Plot.pdf}} \\
\subfloat[]{\includegraphics[width=0.5\textwidth]{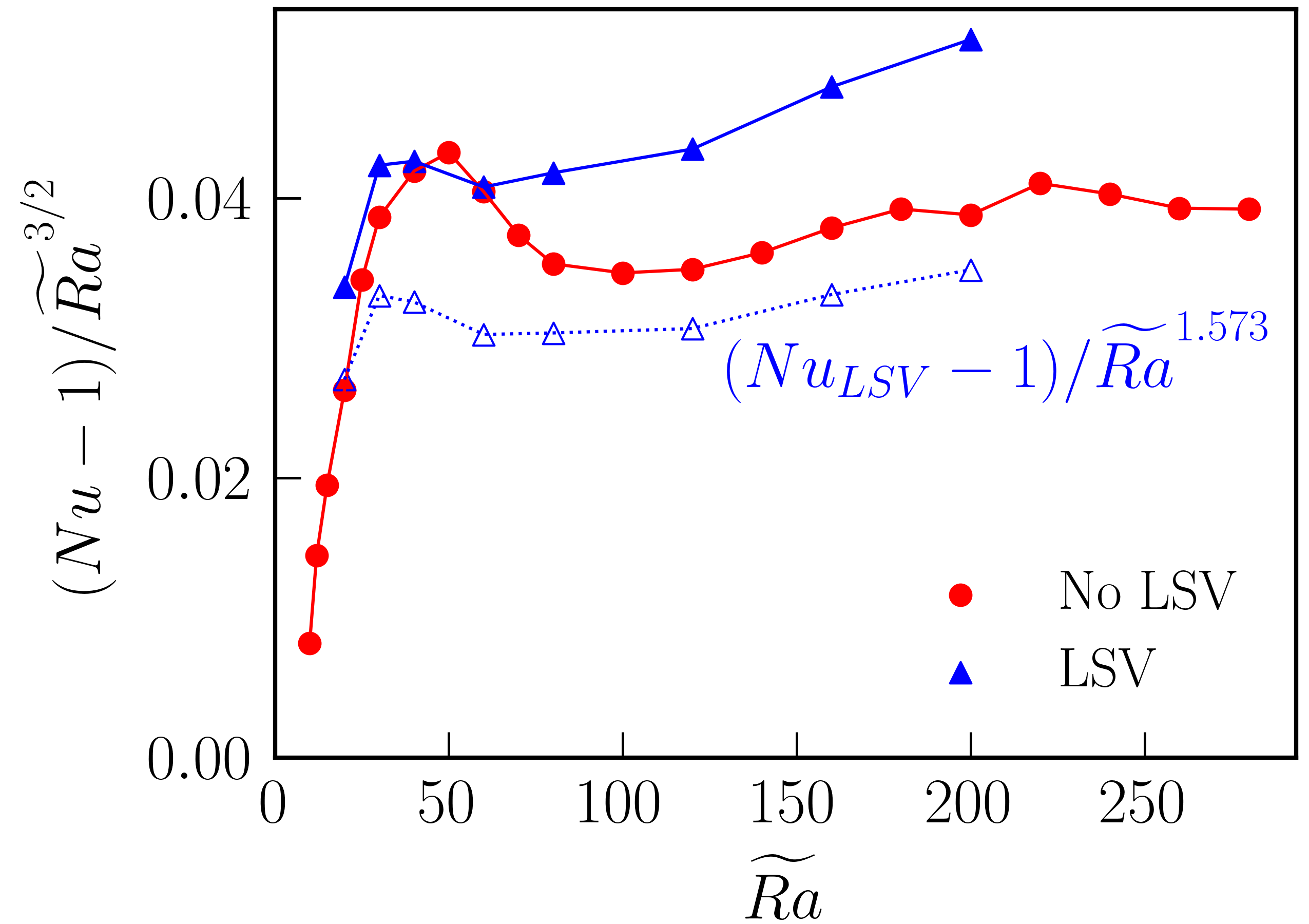}}
	\subfloat[]{\includegraphics[width=0.5\textwidth]{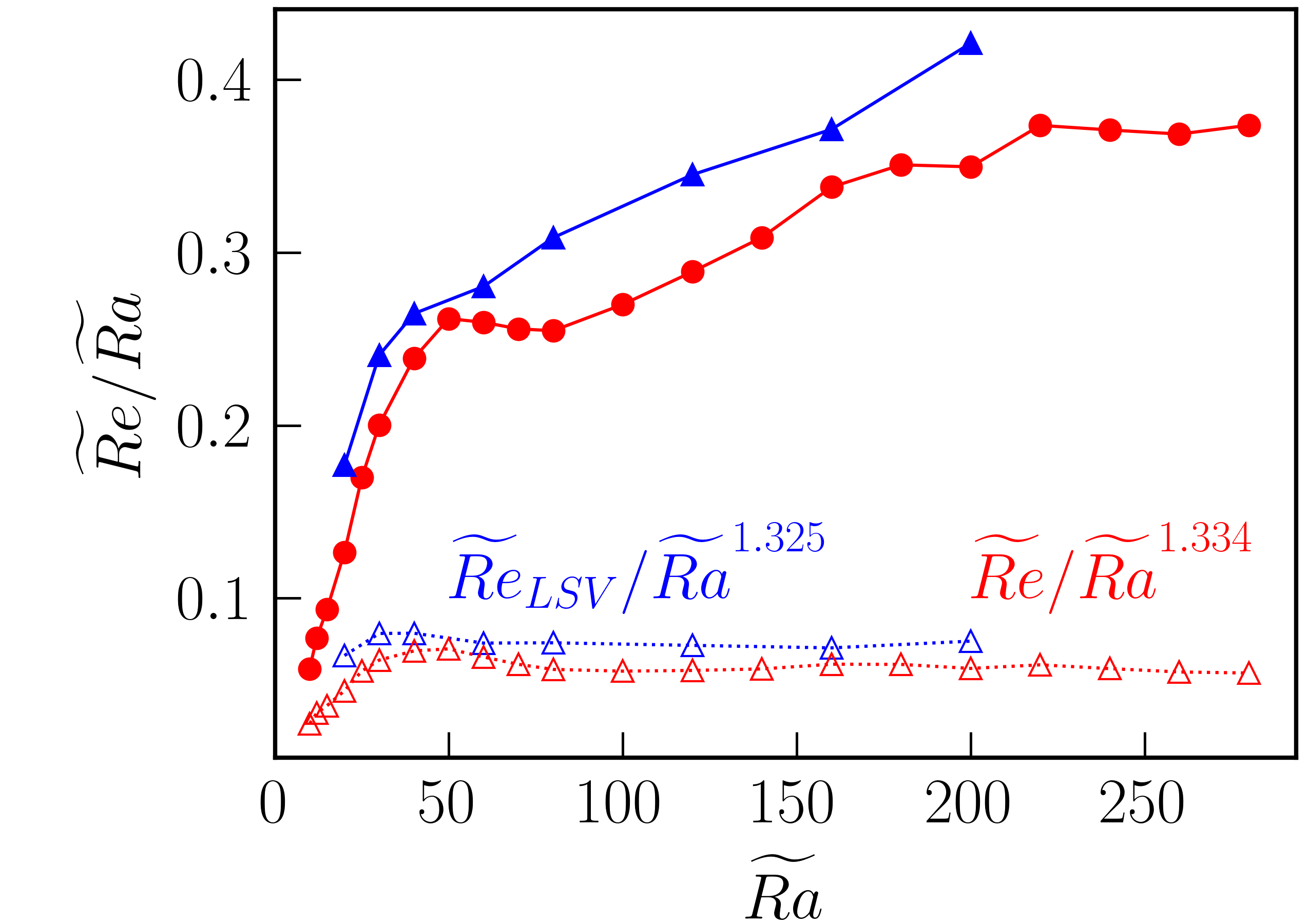}}
	\caption{Heat and momentum transport for all cases: (a) compensated Nusselt number, $Nu/\Rat^{3/2}$; (b) compensated reduced Reynolds number, $\Ret/\Rat$.
		Blue data points are from simulations with a non-zero barotropic component (LSV). Data with open symbols in panels (a) and (b) shows the compensation required to flatten $Nu$ and $\Ret$ for the LSV data. 
	}
	\label{F:Nu_Re}
\end{figure}
Panels (a) and (b) in Figure \ref{F:Nu_Re}  show the compensated values $Nu/\Rat^{3/2}$ and $\Ret/\Rat$, respectively. 
\textcolor{black}{As previously mentioned, these scalings are derived from CIA theory.
For the `No LSV' data we find that both compensated quantities exhibit steep initial increases up to $\Rat = 50$, followed by decreasing values up to $\Rat = 80-100$. The Nusselt number shows a trend that scales somewhat stronger than $Nu \sim \Rat^{3/2}$ in the range $100 < \Rat < 220$.
%\textcolor{red}{KJ: can we say what the $Nu$ exponent is? it scales like the LSV cas, yes, Looks like $\sim 1.834$? TO: I don't know about this... it definitely is not $1.834.$}
 For $\Rat > 220$ we find scaling behavior that may be consistent with the CIA scaling, though the range over which this occurs is admittedly limited. The Reynolds number for the `No LSV' data scales more strongly than the CIA scaling in the range $80 < \Rat<200 $; we find that $\Ret \sim \Rat^{1.334}$ fits the data well in this region of parameter space. For $\Rat > 200$, we find a trend that is consistent with $\Ret \sim \Rat$.
 }
Note that the value of $\Rat \approx 80$ coincides with the point at which the amplitude of the LSV reaches a maximum relative to the small scale convection \citep{sM21}. Interestingly, we find that both the `No LSV' cases and the `LSV' cases exhibit similar scaling trends over a certain range in $\Rat$. We find that the removal of the depth-averaged flow yields reduced heat and momentum transport relative to cases in which this component of the flow is present. 

\begin{figure}
	\begin{center}
		\subfloat[]{\includegraphics[width = 0.33\textwidth]{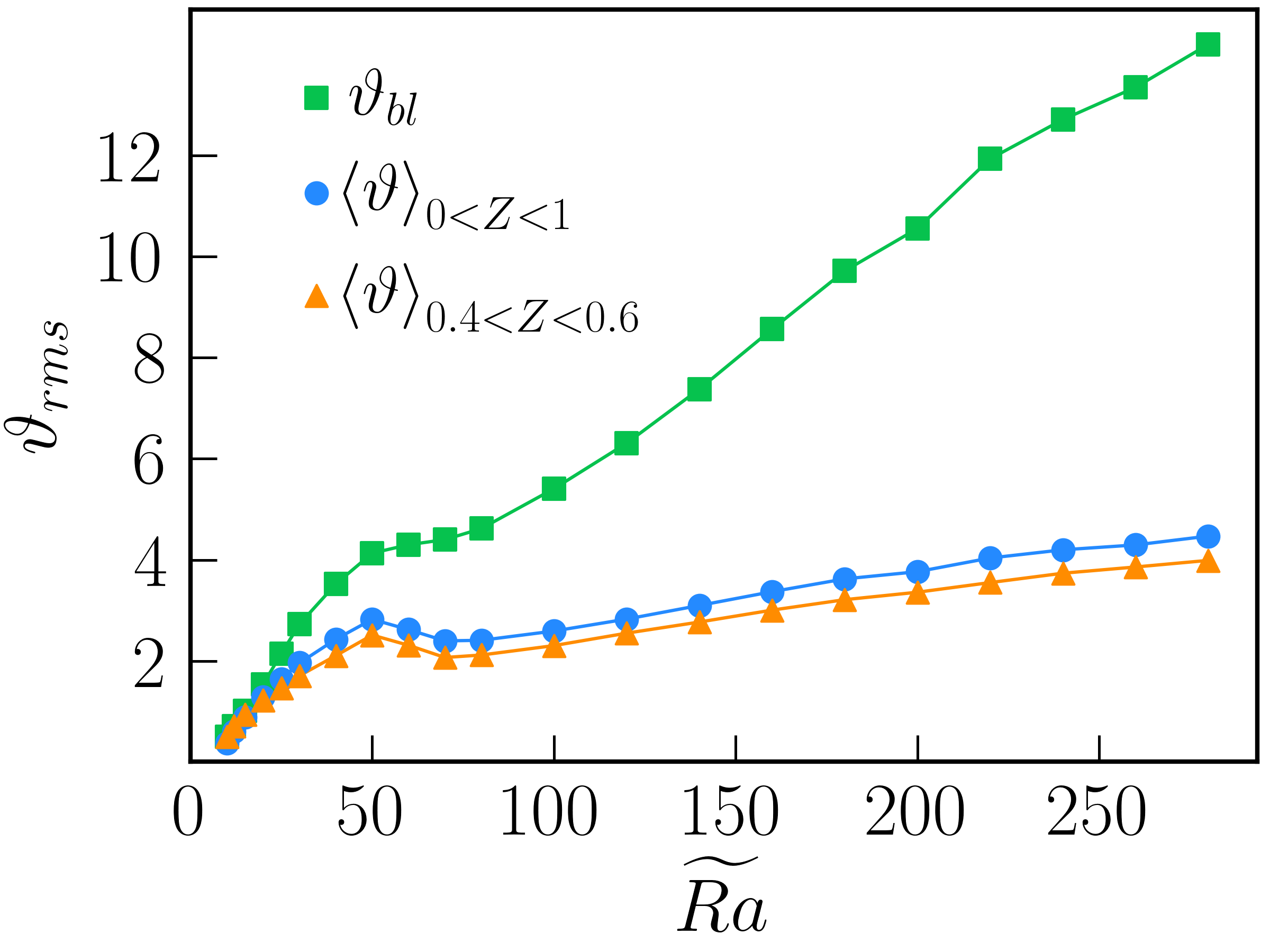}\label{temp_da}} 
		\subfloat[]{\includegraphics[width = 0.33\textwidth]{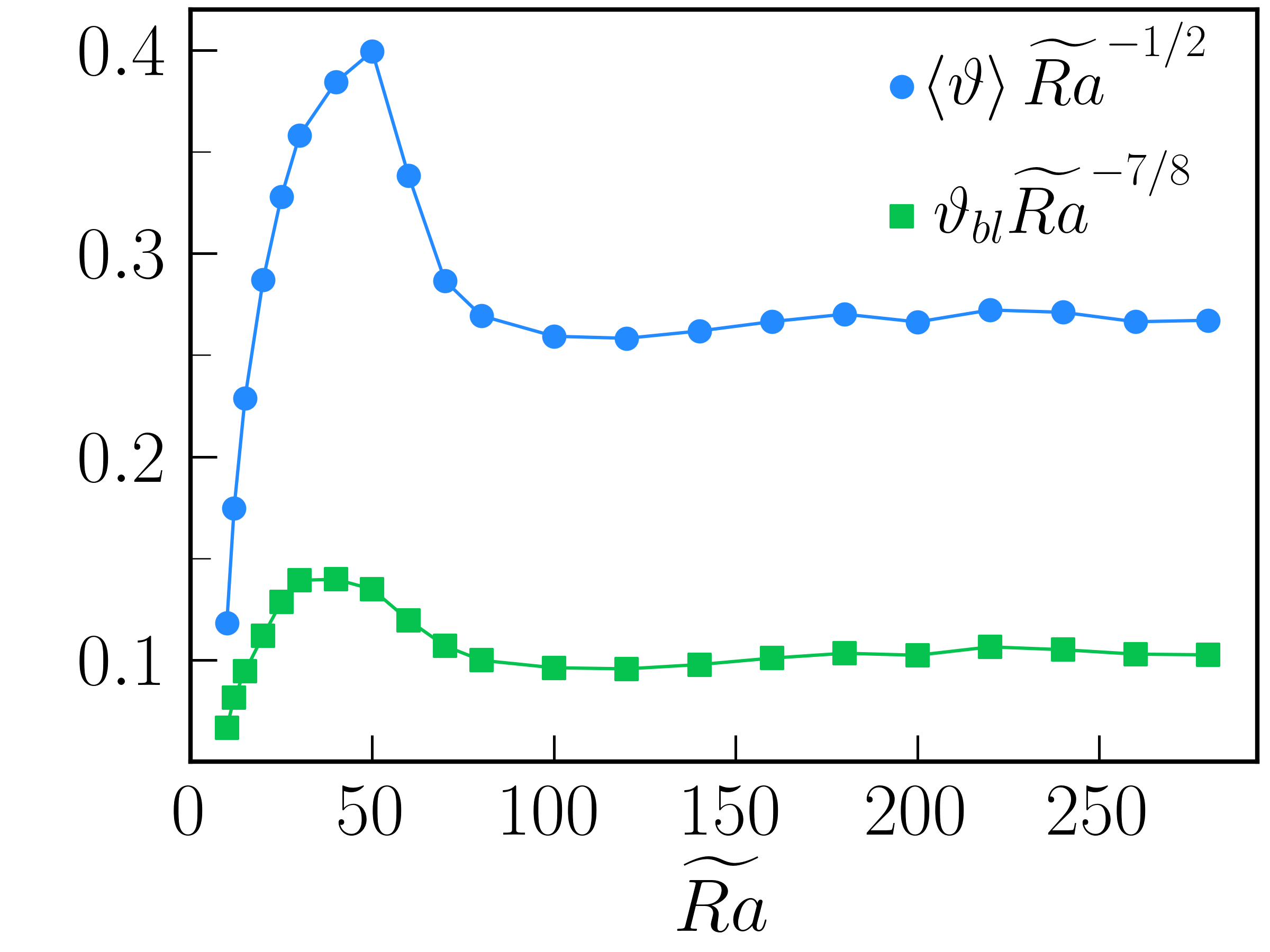}\label{temp_max_da}} 
       	\subfloat[]{\includegraphics[width=0.33\textwidth]{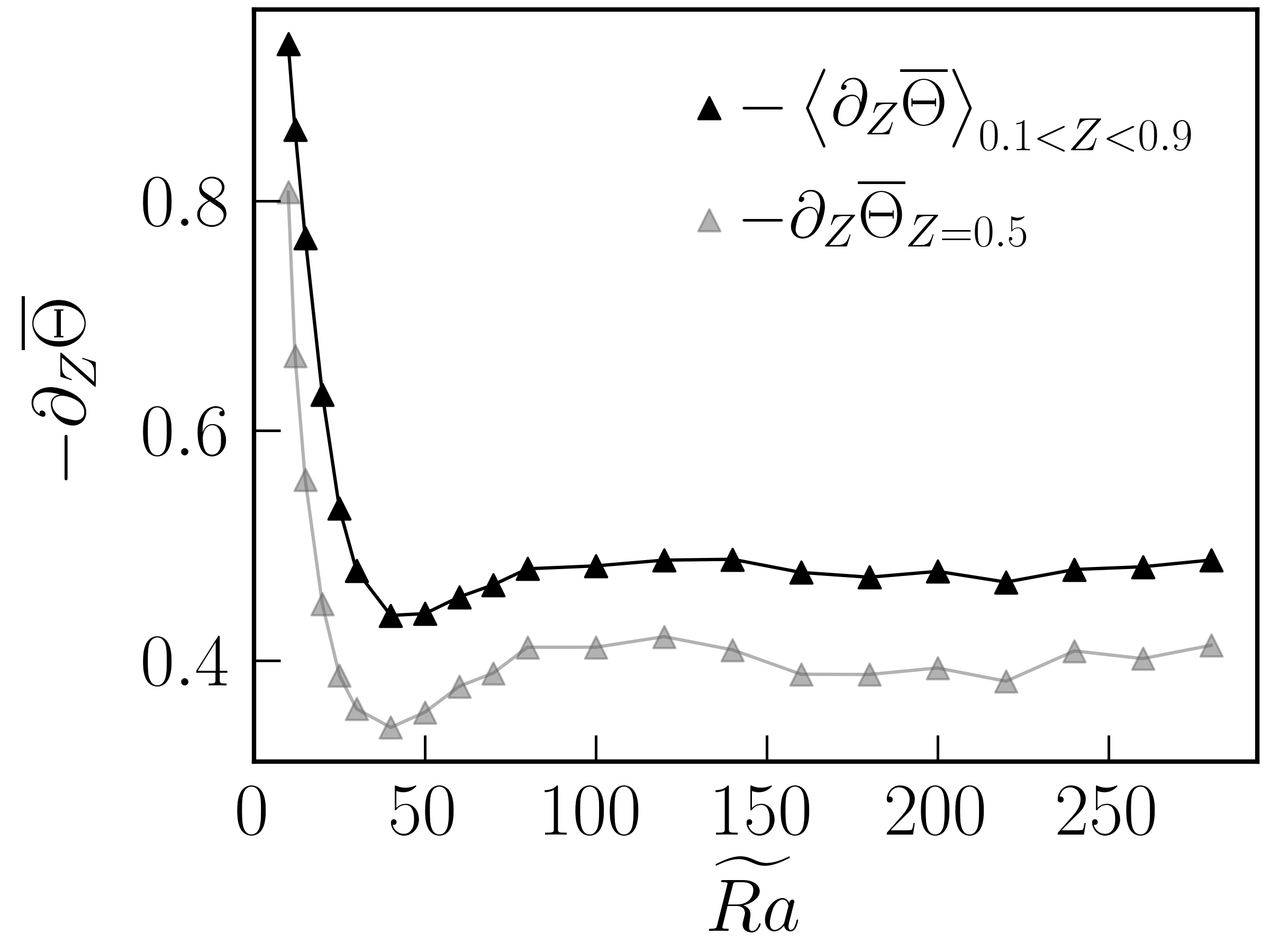}\label{dz_temp}}
	\end{center}
	\caption{\protect\subref{temp_da} Various measures of the rms fluctuating temperature vs. $\Rat$;
		\protect\subref{temp_max_da} compensated fluctuating temperature;
		%The results for $\vartheta_{bl}$ are consistent with a $\Rat^{7/8}$, scaling but do not rule out nearby fractional powers such as $3/4$ or $1$. A powerfit for $\Rat > 80$ gives $\vartheta_{rms}\sim\Rat^{0.776}$.
\protect\subref{dz_temp} mean temperature gradient vs. $\Rat$. The vertical averaging range is indicated with subscripts in (a) and (c).
%In \protect\subref{temp_da}, \protect\subref{temp_max_da}, \ding{110} are computed at the edge of the thermal boundary layer, defined as the location of max($\vartheta_{rms}$), blue \ding{108} are averaged over the entire depth and orange \ding{115} are averaged over the mid-depth range $0.4 < Z < 0.6$.
}
\label{F:temp}
\end{figure}

With the scalings of $Nu \sim \Rat^{3/2}$ and $\Ret \sim \Rat$, equations \eqref{E:Nu} and \eqref{E:Re} indicate that the fluctuating temperature scales as $\vartheta \sim \Rat^{1/2}$. 
 %and the nondimensional convective velocity scales according to rotational free-fall $U^{*}\sim g\alpha \Delta T/2\Omega$. 
To test this scaling we compute rms values of the fluctuating temperature over various depths of the flow domain: (1) the maximum value taken at the `edge', or inflection point, of the thermal boundary layer; (2) averaged over the entire layer depth $0 \le Z \le 1$; and (3) averaged over the mid-depth range, $0.4 < Z < 0.6$. 
%Figure \ref{figure:temp_rms}(a) shows vertical profiles of the rms temperature fluctuation for a selection of $\Rat$ values. The rescaled data in Figure \ref{figure:temp_rms}(b) shows that the data collapses well with this rescaling. In particular, the profiles show good collapse for $\Rat > 80$ which is consistent with the observed trends for $Nu$ and $\Ret$ given in Figures \ref{F:Nu_Re}(c) and (d), respectively, that indicate both (compensated) quantities are nearly constant for Rayleigh numbers larger than this value. 
Figure \ref{F:temp}(a) shows these values; Figure \ref{F:temp}(b) shows the compensated values for the thermal boundary layer and entire depth. 
We find that both the mid-depth and total depth rms values scale with $\Rat$ in the same manner, whereas the thermal boundary layer shows a stronger dependence on $\Rat$ (Figure \ref{F:temp}\protect\subref{temp_max_da}), suggesting that the relative thinness of the thermal boundary layer translates to an overall weaker influence of this region on the observed scaling of the Nusselt number. 
%\textcolor{black}{The $\vartheta \sim \Rat^{1/2}$ scaling agrees with the CIA prediction, but only agrees with our empirical scalings for $\Rat>200$.} 
From the data, it was difficult to determine the powerlaw scaling for $\vartheta_{rms}$ in the boundary layer, as results differed significantly depending on the range of data used.
For example, fitting the data in Figure \ref{F:temp}\protect\subref{temp_max_da} for $\widetilde{Ra}>80$ gives a \textcolor{black}{$\Rat^{0.95}$} relation while fitting over only the last three points ($\Rat>200$) yields a \textcolor{black}{$\Rat^{0.70}$} scaling. 
Nevertheless, we conclude that our results for $\vartheta_{rms}$ in the thermal boundary layer are consistent with the theoretically deduced $\Rat^{7/8}$ scaling discussed in \textcolor{black}{Ref.~\cite{kJ12b}.}
As previously reported in \textcolor{black}{Ref.~\citep{mS06}}, we observe saturation of the interior mean temperature gradient with a value of $-\partial_{Z}\overline\Theta \approx 0.4$-$0.5$ (Figure \ref{F:temp}\protect\subref{dz_temp}).

\subsection{Length scales and flow structure}
\label{sec:ls}

\begin{figure}
\begin{center}
	\subfloat[]{\includegraphics[width=0.49\textwidth]{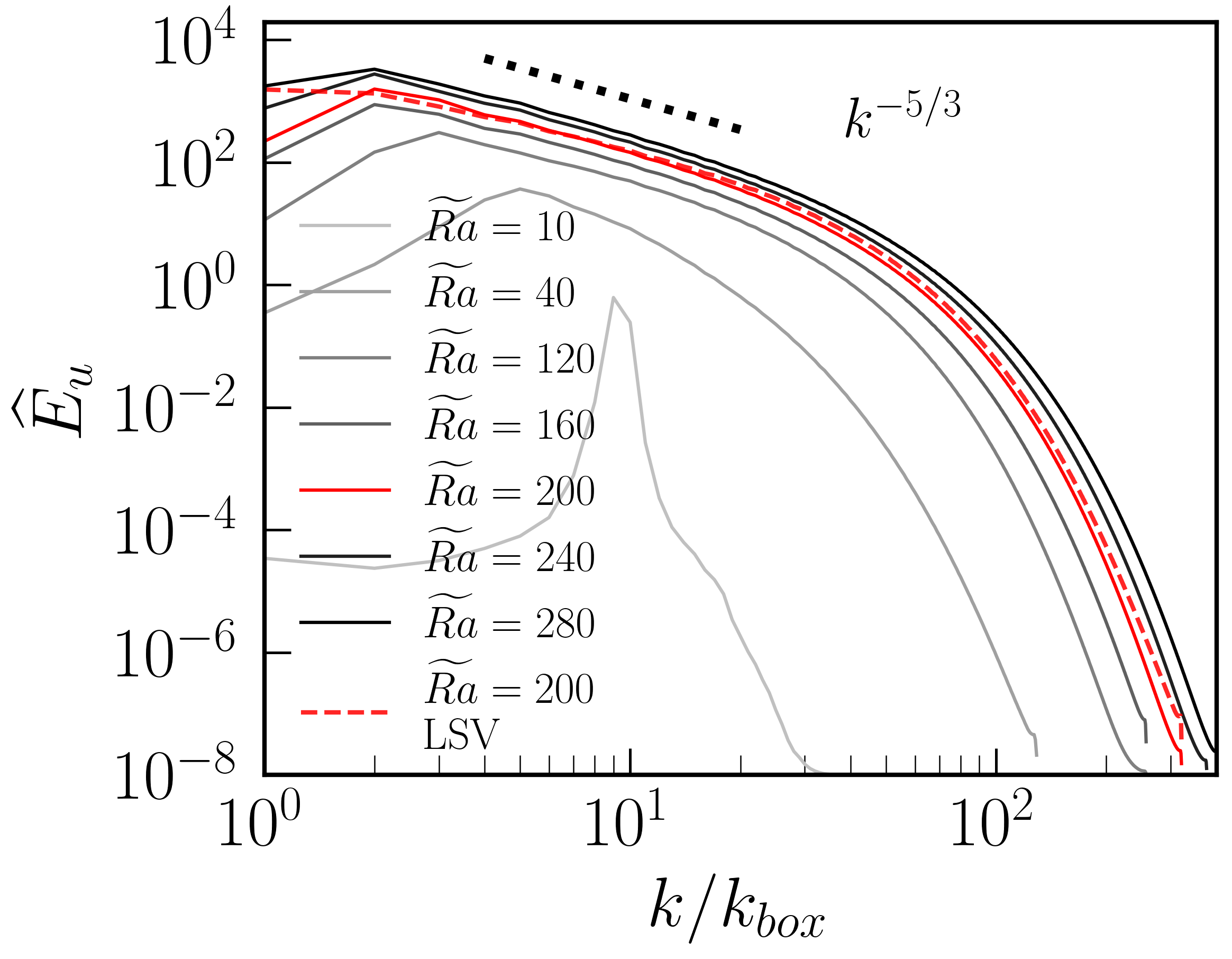}\label{energy_spectra}} 
	\subfloat[]{\includegraphics[width=0.49\textwidth]{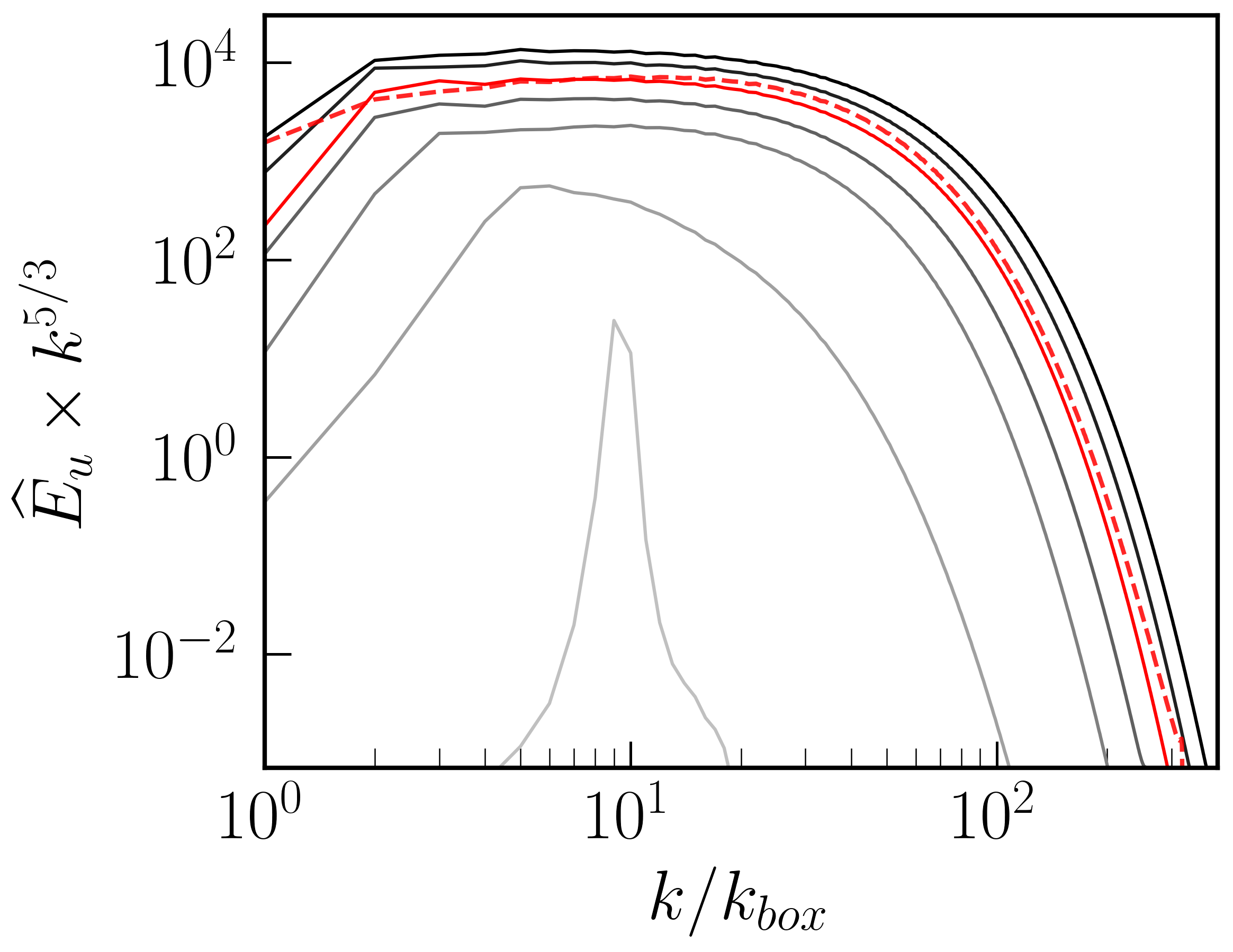}\label{energy_spectra_scaled}}\\
	\subfloat[]{\includegraphics[width =0.49\textwidth]{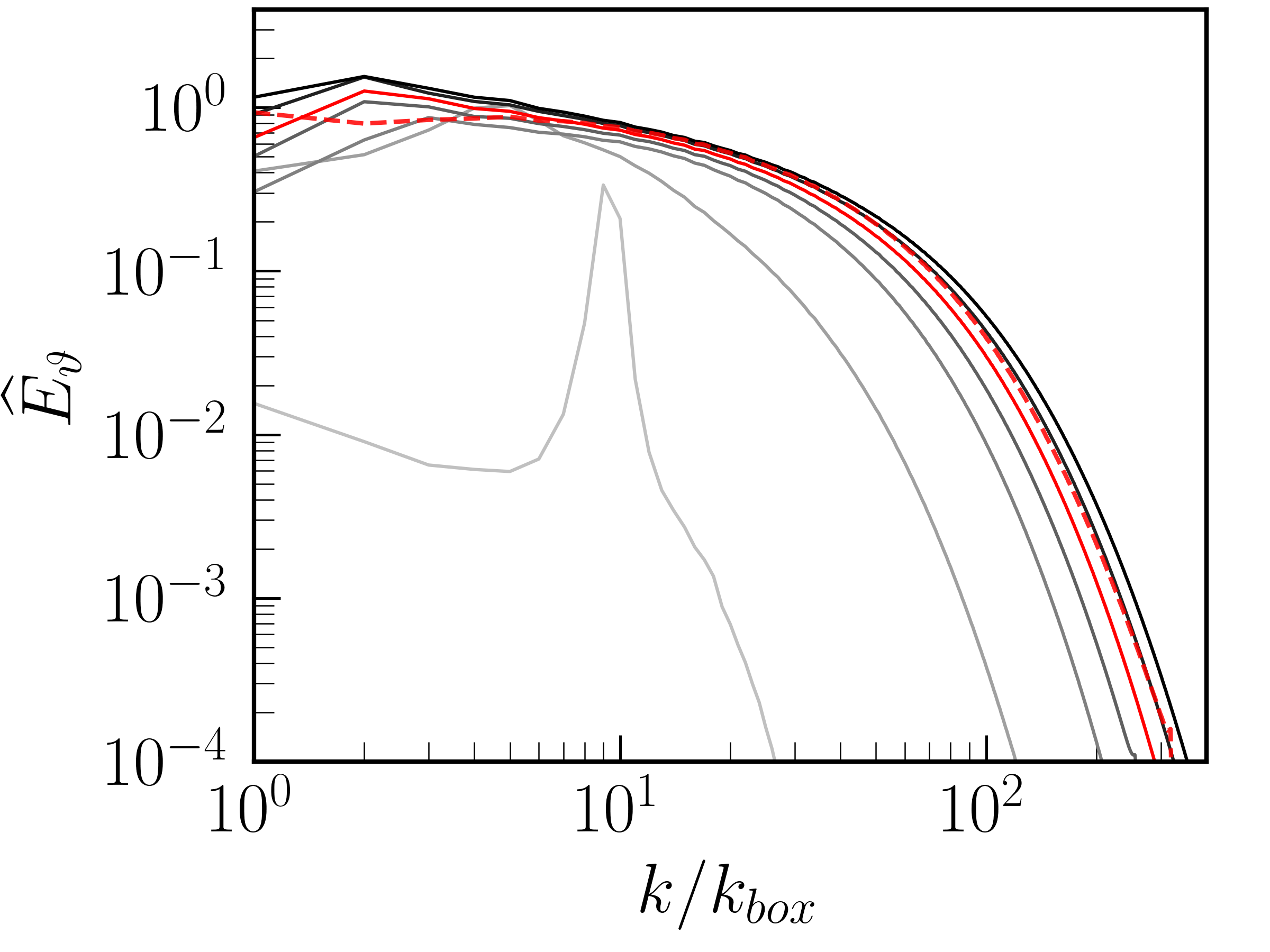}\label{temp_spectra}} 
	\subfloat[]{\includegraphics[width =0.49\textwidth]{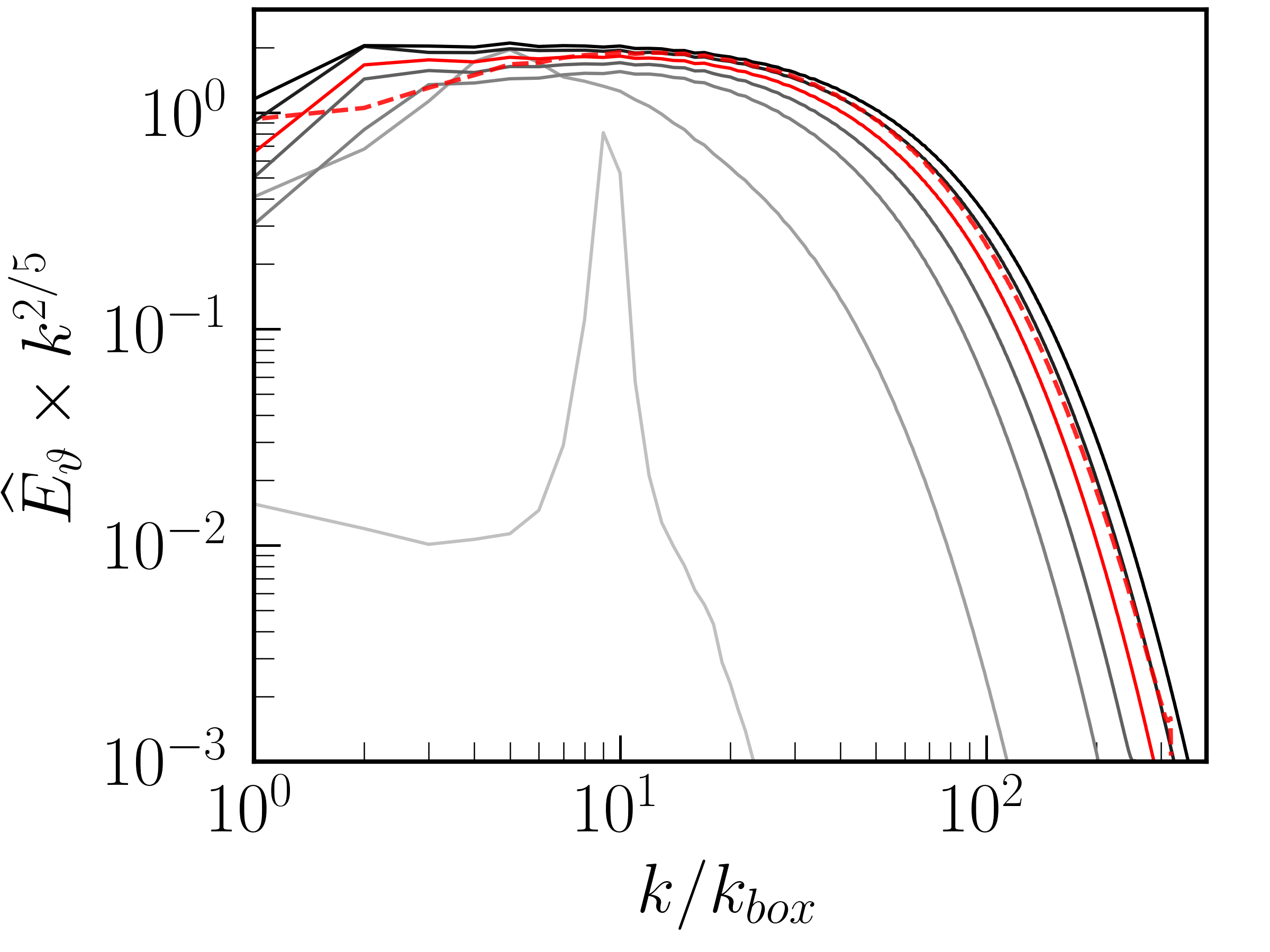}\label{temp_spectra_scaled}}
\end{center}
	\caption{Spectra for select values of $\Rat$: (a) kinetic energy spectra vs. box normalized wavenumber; (b) compensated kinetic energy spectra; (c) temperature variance spectra; (d) compensated temperature variance spectra.
Red dashed curves in show spectra for $\Rat=200$ with the LSV.
%The energy spectra does not follow a $-5/3$ power law region, suggesting the absence of a traditional inertial subrange. 
%\protect\subref{lengthscales} Length scales vs. $\Rat $; $\ell_{I}$ is the integral scale, $\ell_{T}$ is the Taylor microscale, $\ell_{K}$ is the Kolmogorov scale, and $\ell_{I}^{\vartheta}$ is the integral scale for the fluctuating temperature. 
%\textcolor{black}{The widehat in the axis label of (b) should not go over the subscript. More padding for labels. Let's add compensated temperature variance spectra in (d) using whatever scaling is necessary to flatten the midrange.} \textcolor{red}{KJ: In plot (a) plot $k^{-5/2}$ and $k^{-5/3}$ at a guide.  in plot (c) plot $k^{-2/5}$.}
}
\label{F:spectra}
\end{figure}

Kinetic energy spectra, $\widehat{E}_u(k)$, and temperature variance spectra, $\widehat{E}_{\vartheta}(k)$, are shown in Figure \ref{F:spectra} for a range of Rayleigh numbers. 
%The horizontal wavenumber is denoted by $\mathbf{k} = (k_x, k_y)$ and the modulus is $k = |\mathbf{k}| = \sqrt{k_x^2 + k_y^2}$. 
As expected, near the onset of convection for $\Rat = 10$, both $\widehat{E}_u(k)$ and $\widehat{E}_{\vartheta}(k)$ show a well-defined peak near $k/k_{box}=10$, corresponding to ten unstable wavelengths in the domain.
A broadening of dynamically active wavenumbers is observed in both spectra as $\Rat$ is increased. 
%There is a trend to a self-similar structure for the largest values of $\Rat$.
Figure \ref{F:spectra}\subref{energy_spectra_scaled} shows the kinetic energy spectra compensated by the Kolmogorov scaling of $k^{-5/3}$, 
\textcolor{black}{ which we find agrees well with the data.} 
The red dashed curve shows the baroclinic kinetic energy spectra for a simulation with $\Rat = 200$ that includes the LSV. 
Comparison of the spectra at $\Rat = 200$ for cases with and without the LSV indicates that significantly more energy is present in the smallest wavenumbers when the LSV is present. 
This difference, along with the findings reported in Section \ref{S:Nu_Re}, suggests that the presence of more energy in the largest length scales contributes to more efficient heat and momentum transport.
%This difference indicates that the presence of a large scale overturning circulation for the cases with the LSV leads to more efficient heat and momentum transport  \textcolor{red}{KJ: while true, I think we can only make this statement if we are observing the vertical KE spectra}.
The temperature variance spectra shown in Figure \ref{F:spectra}\subref{temp_spectra} corroborate this behavior; with the LSV we find significantly more energy in the $k/k_{box}=1$ mode in comparison to the same case without a LSV. 
We also note that in comparison to $\widehat{E}_{u}(k)$, we find that $\widehat{E}_{\vartheta}(k)$ shows a significantly slower decay in amplitude with increasing $k$, indicating that buoyancy forcing in the vicinity of the critical wavenumber remains important even at very large values of $\Rat$. The compensated temperature variance spectra shown in Figure \ref{F:spectra}(d) illustrates this slow decay where an empirical scaling of $k^{-2/5}$ flattens the spectra in the intermediate wavenumber range.

%For a small range of intermediate wavenumber ($4\leq k \leq 10$ for $\Rat = 200$), the kinetic energy is slightly greater for the non-LSV case.
%This behavior is also reflected in the temperature spectra, although less markedly and for a smaller range of wavenumber ($3\leq k\leq 4$ for $\Rat = 200$).
%This indicates that the LSV reduces the energy that is present in a small range of $O(1)$ modes larger than the box-filling mode (i.e.~$k=1$). \textcolor{red}{MAC: not sure if this is significant -- the difference is so small that it could be time averaging?}

\begin{figure}
\begin{center}
	\subfloat[]{\includegraphics[width =0.5\textwidth]{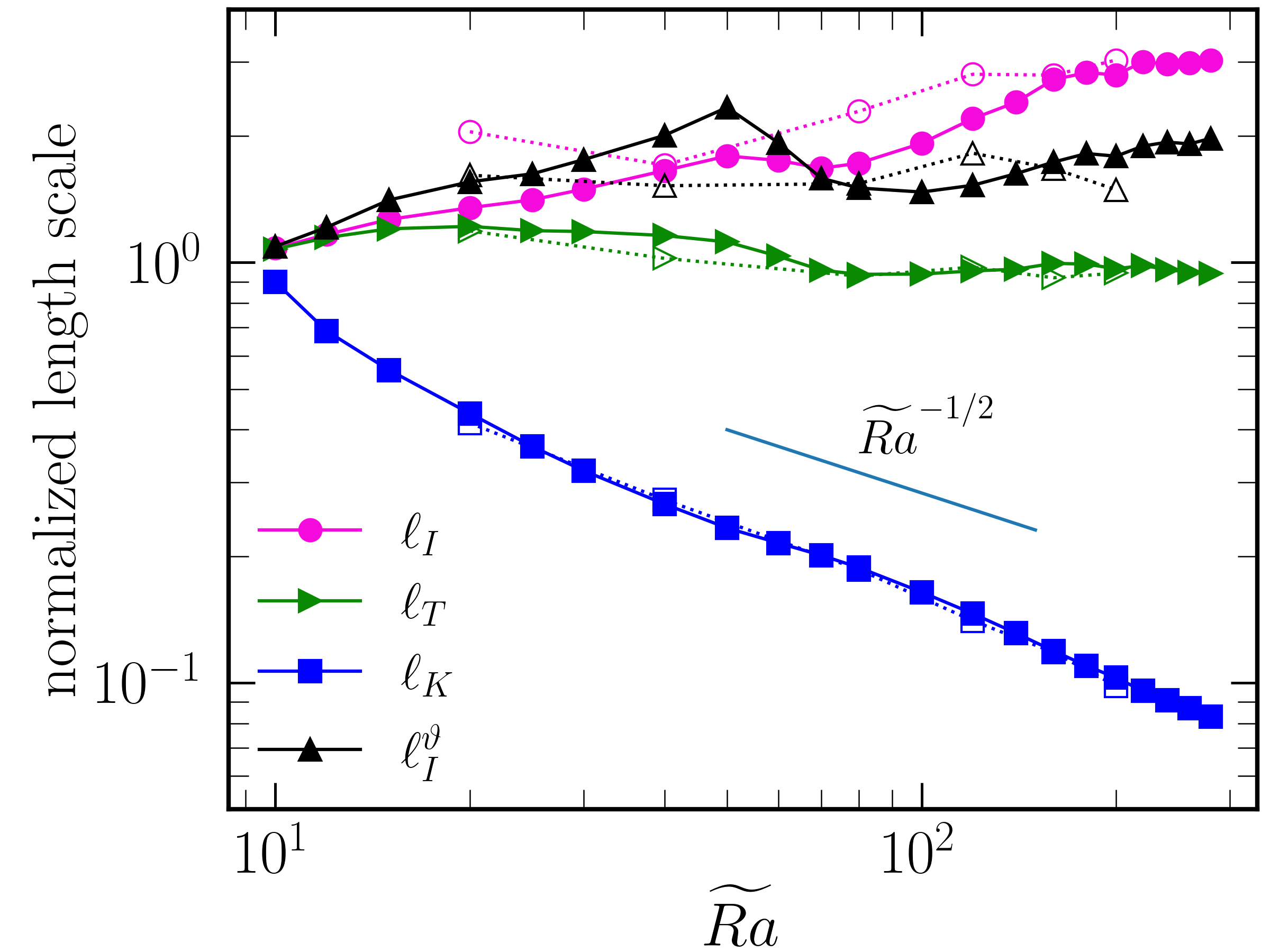}\label{lengthscales}}
	\subfloat[]{\includegraphics[width =0.5\textwidth]{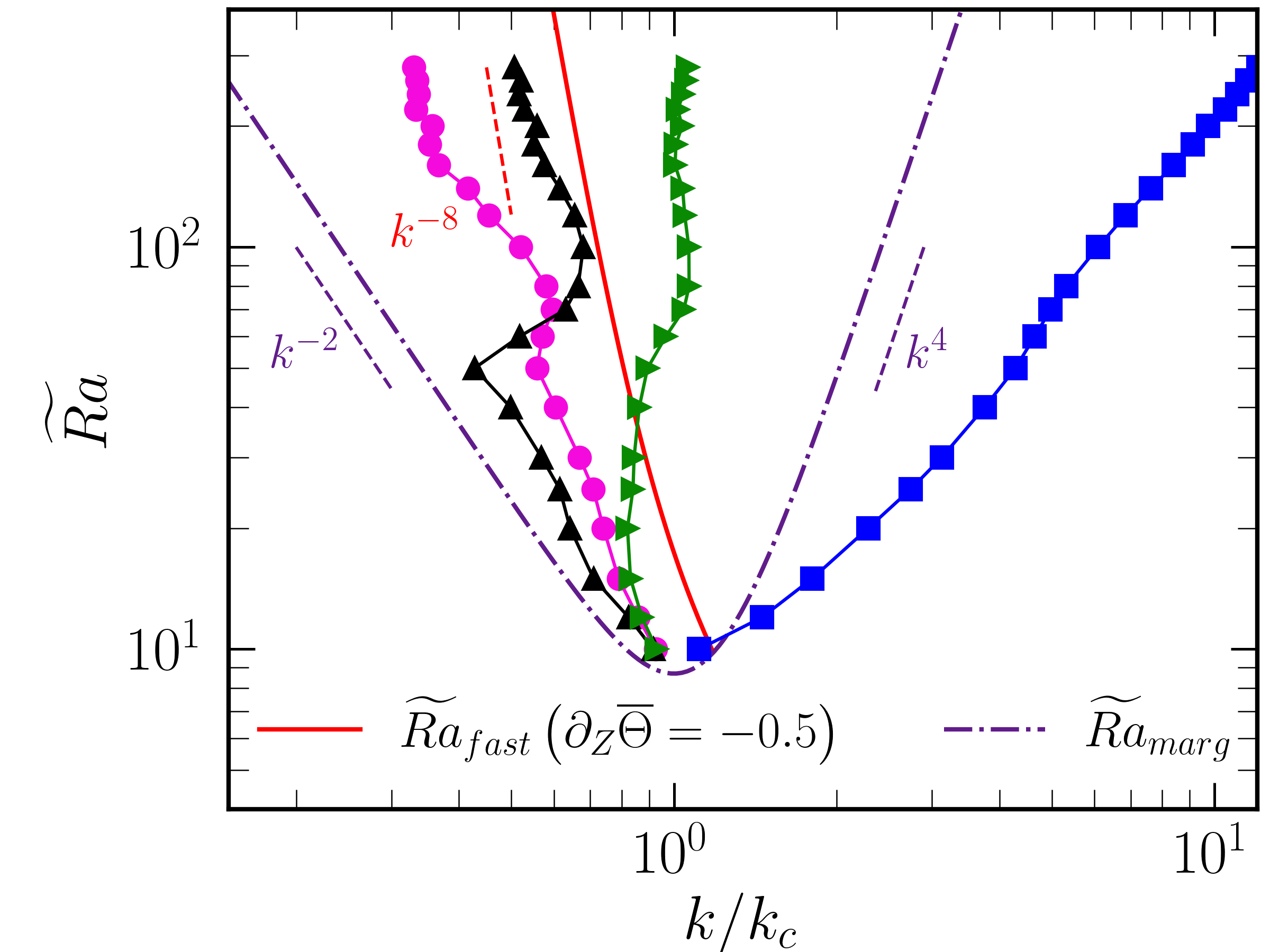}\label{fastest_mode}}
\end{center}
	\caption{(a) Length scales versus $\Rat$ for all simulations where $\ell_{I}$ is the integral scale, $\ell_{T}$ is the Taylor microscale, $\ell_{K}$ is the Kolmogorov scale,  and $\ell_{I}^{\vartheta}$ is the integral scale for the fluctuating temperature.
In (a) open symbols correspond to LSV cases and the light blue line shows the $\Rat^{-1/2}$ scaling for reference.
(b) Comparison of computed length scales with linear stability theory.
$\Rat_{marg}$ is the marginal stability curve. $\Rat_{fast}$ is the fastest growing mode in the linear stability analysis about the saturated mean temperature profile. For large $\Rat$, $\Rat_{fast}\sim k^{-8}$. The critical wavelength and critical wavenumber are denoted by $(\ell_c, k_c) = (4.8154, 1.3048)$.
}
\label{F:LS}
\end{figure}

%\begin{figure}
%\begin{center}
%	%\subfloat[]{\includegraphics[width =0.48\textwidth]{LengthScalesPlot_Scaled.pdf}\label{lengthscales_scaled}}
%\end{center}
%\caption{Length scales vs. $\Rat$; $\ell_{I}$ is the integral scale and $\ell_{T}$ is the Taylor microscale; $ell_{K}$ is the Kolmogorov scale. Unfilled points correspond to LSV cases and the light blue line plots a $\Rat^{1/2}$ scaling.}
%\label{F:lengthscales}
%\end{figure}
Various length scales of the velocity field can be computed from the kinetic energy spectra. Definitions for the integral length scale, the Taylor microscale and the Kolmogorov length scale in terms of the kinetic energy spectrum and vorticity (dissipation) are given in equation \eqref{eqn:scales}.
Each of these quantities are plotted in Figure \ref{F:LS}\subref{lengthscales} for all simulations. We also plot the integral scale $\ell^\vartheta_I$ computed from  $\widehat {E}_{\vartheta}$, which represents a measure of the buoyancy driving scale. For the LSV cases, the length scales are calculated using only the baroclinic kinetic energy for better comparison with the present results.
The length scales are shown in units of the critical horizontal wavelength and we find that all length scales are close to this critical wavelength just above the onset of convection (i.e.~when $\Rat = 10$). 
Generally, the integral length scale is found to be a slowly increasing function of $\Rat$ regardless of whether the LSV is present or not. 
\textcolor{black}{At $\Rat = 280$ we find that the integral scale is approximately three times the size of the critical wavelength.}
%In comparison to cases without the LSV, the LSV cases show a slightly faster increase of $\ell_I$ for $\Rat > 40$. 
%At $\Rat = 280$ we find that the integral scale is approximately twice the size of the critical wavelength. 
In comparison to the integral scale, the Taylor microscale shows an initial increase up to \textcolor{black}{$\Rat=20$ followed by a slight decrease}. For $\Rat \geq 80,$ the average value of the Taylor microscale is \textcolor{black}{$\ell_{T} = 0.96$} and the standard deviation is $0.02$.
Both the integral length scale and the Taylor microscale exhibit scaling behavior that is weaker than the $\Rat^{1/2}$ scaling law. 
The Kolmogorov scale exhibits an obvious decrease with $\Rat$ and, for the parameter regime accessible here, is the only computed length scale that becomes significantly different in value than the critical wavelength. For LSV cases, $\ell_{K}$ is found to be slightly smaller in comparison to cases without the LSV; this result is in agreement with the heat transport data which shows LSV cases have larger heat transport and therefore larger viscous dissipation. 
The integral temperature scale remains $O\lb 1 \rb$ up to $\Rat = 280$, which suggests that the buoyancy forcing scale occurs at a viscous length scale.
\textcolor{black}{$\ist$ reaches a maximum value at $\Rat = 50,$ which is coincident with the local maxima for the scaled $Nu$ and $\Ret$ data shown in figure \ref{F:Nu_Re}.}
Around $\Rat \geq 120$, $\ist$ begins to slowly increase. 
This behaviour is co-incident with a decrease in $\ist$ for the LSV cases. 
It appears that a crossover in $\ist$ for cases with the LSV and cases without a LSV occurs around \textcolor{black}{$\Rat = 180$}. 
This result suggests that at large $\Rat$, the LSV facilitates the buoyancy forcing at smaller scales than those without the LSV.

%\begin{figure}
%\begin{center}
%  \includegraphics[width=0.75\textwidth]{Ra_vs_k.png}
%\end{center}
%	\caption{\textcolor{red}{Work in progress here. Do we want this plot or something similar?}}
%\label{F:Ra_vs_k}
%\end{figure}

%\textcolor{black}{In Figure \ref{F:Ra_vs_k} we compare the computed length scales with the marginal stability curve, defined by
%\be
%\Rat_{marg} = k^4 + \pi^2 k^{-2}. 
%\ee
%Also shown is linearly unstable curve that has the largest growth rate. For the marginal stability curve, the two limiting scalings for small wavenumber and large wavenumber are also indicated as $\sim k^{-2}$ and $k^4$, respectively. Importantly, the small wavenumber scaling yields an unstable wavelength of
%%\be
%$\lambda_{long} \propto \Rat^{1/2}$ coinciding with inviscid CIA integral scale
%%\ee
%and the fastest growing mode a wavelength of  $\lambda_{fast} \propto \Rat^{1/8}$.
%These results suggest that the integral length scale is consistent with the fastest growing wavelength in the system.}

Due to the similarity between the computed length scales and the linearly unstable wavelength, we show an alternative presentation of length scales based on comparisons with the marginal stability boundary and the fastest growing linearly unstable modes in Figure \ref{F:LS}\subref{fastest_mode}. The marginal stability boundary is defined by
\be
\Rat_{marg} = k^4 + \pi^2 k^{-2}. 
\ee
The small and large wavenumber scalings of $k^{-2}$ and $k^4$ are also shown where we note that the former  is consistent with CIA theory in that it represents a diffusion-free scaling behavior. We find that none of the computed length scales exhibit scaling behavior that is similar to this diffusion-free trend, \textcolor{black}{though there are ranges of $\Rat$ over which these length scales grow faster than the diffusion-free trend.}
%For comparison, the asymptotic scalings in the large $\Rat$ limits are also highlighted. 
%that have a wavelength of $\lambda_{fast} \propto \Rat^{1/8}$ (In a limit?). 
%We observe a similarly slow scaling of the integral length scale with increasing $\Rat$ as opposed to the $~k^{-2}$ prediction by CIA theory.

For any constant (non-zero) mean temperature gradient, a linear stability analysis can be performed to acquire a growth rate $\lambda\lb k,\Rat\rb. $ 
\textcolor{black}{We perform this analysis with the linear eigenfunctions
	\[\psi = \Psi_{k_x,k_y}e^{\lambda t}e^{i\lb k_x x + k_y y\rb }\cos \lb \pi Z\rb, \]
\[\left[ w,\vartheta \right] = \left[ W_{k_x,k_y}, \theta_{k_{x},k_{y}}\right] e^{\lambda t}e^{i\lb k_x x + k_y y\rb }\sin \lb \pi Z\rb, \]
which can be plugged into equations \eqref{eqn:barotropicvort}-\eqref{eqn:barotropicdff} using the relation $\nabla_{\perp}^{2}\psi = \zeta.$ Since the integral scale appears to be a viscous length scale, we believe that this choice is somewhat justified. }
The fastest growing mode is found by taking the global maximum of $\lambda$ for a given value of $\Rat.$
In the limit of large $\Rat$, this analysis suggests that the length scale associated with the fastest growing mode scales like $\ell_{fast}\sim \Rat^{-1/8}$; we find that $\ell_I$ shows a scaling behavior that is similar to this trend. \textcolor{black}{Figure \ref{fastest_mode} shows the fastest growing mode for $\partial_{Z}\overline\Theta = 0.5$. However, the scaling behavior of the fastest growing mode does not vary with different values of $\partial_{Z}\overline\Theta.$}

%we compare the computed length scales with the marginal stability curve, defined by
%\be
%\Rat_{marg} = k^4 + \pi^2 k^{-2}. 
%\ee
%Also shown is linearly unstable curve that has the largest growth rate. For the marginal stability curve, the two limiting scalings for small wavenumber and large wavenumber are also indicated as $\sim k^{-2}$ and $k^4$, respectively. Importantly, the small wavenumber scaling yields an unstable wavelength of
%%\be
%$\lambda_{long} \propto \Rat^{1/2}$ coinciding with inviscid CIA integral scale
%%\ee
%and the fastest growing mode a wavelength of  $\lambda_{fast} \propto \Rat^{1/8}$.
%These results suggest that the integral length scale is consistent with the fastest growing wavelength in the system.}

\begin{figure}
	\begin{center}
		\subfloat[]{\includegraphics[width=0.33\textwidth]{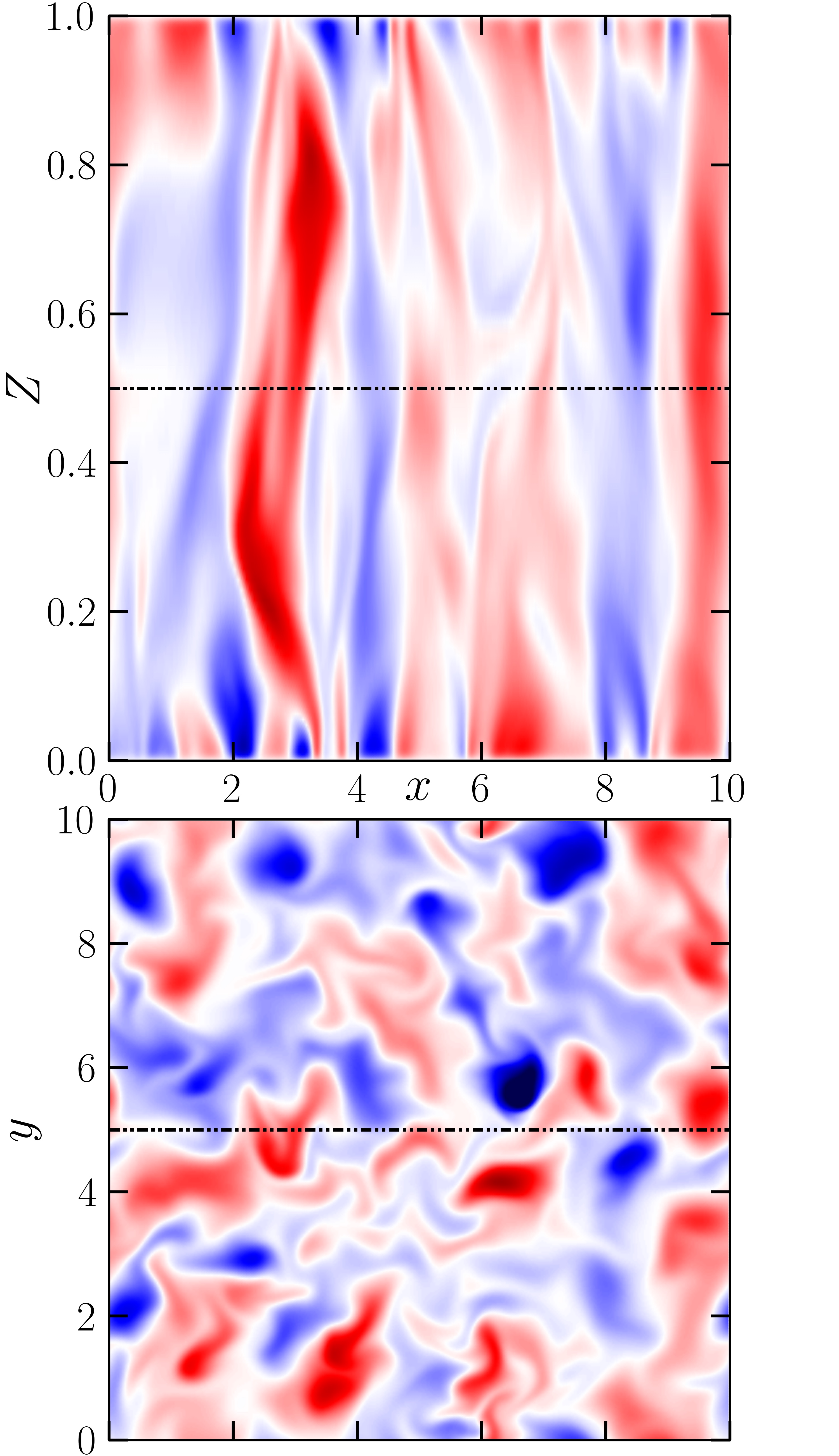}}
		%\subfloat[]{\includegraphics[width=0.53\textwidth]{R100temp_midplane_wscale.png}} \\
		\subfloat[]{\includegraphics[width=0.33\textwidth]{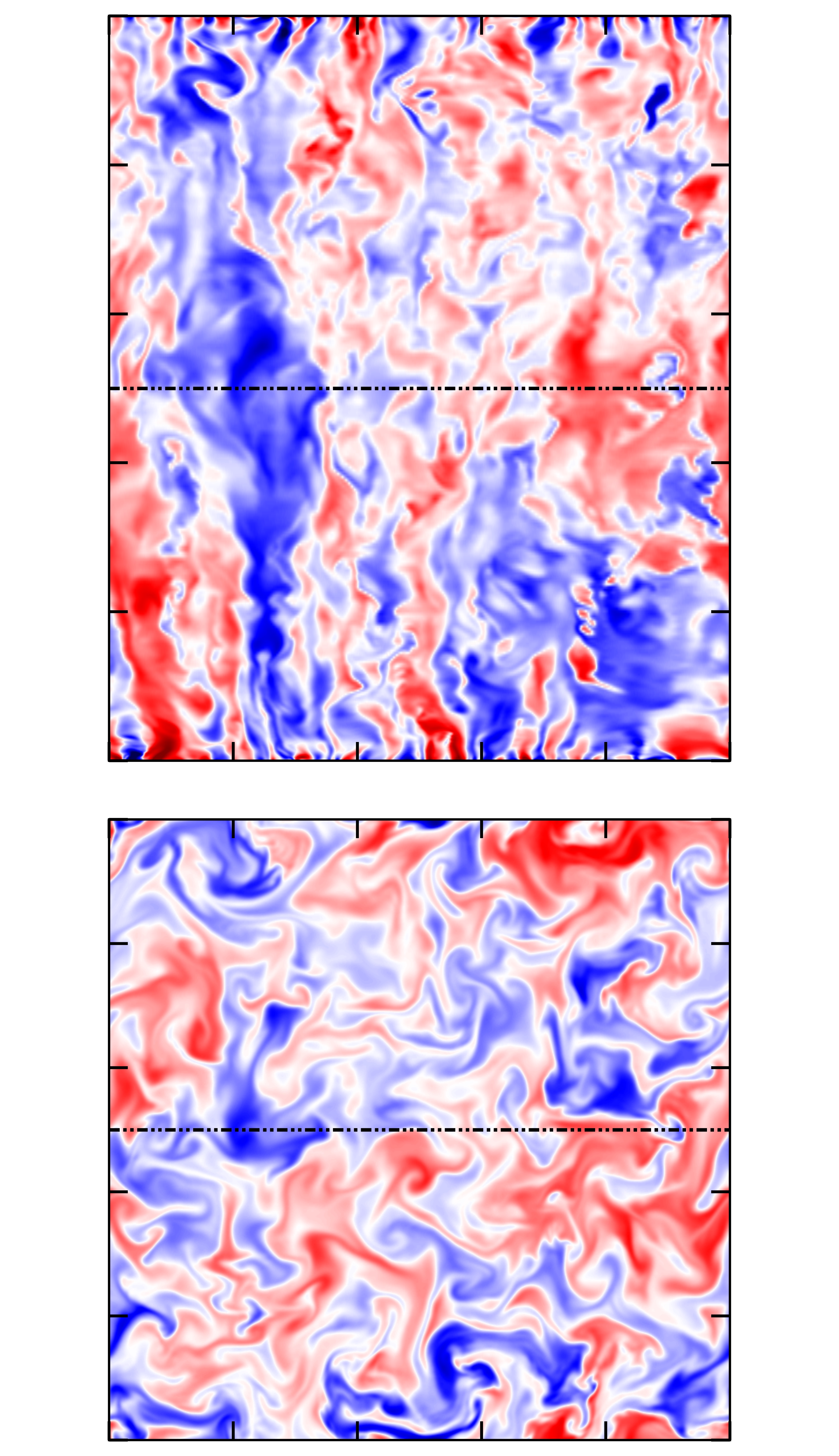}}
		\subfloat[]{\includegraphics[width=0.33\textwidth]{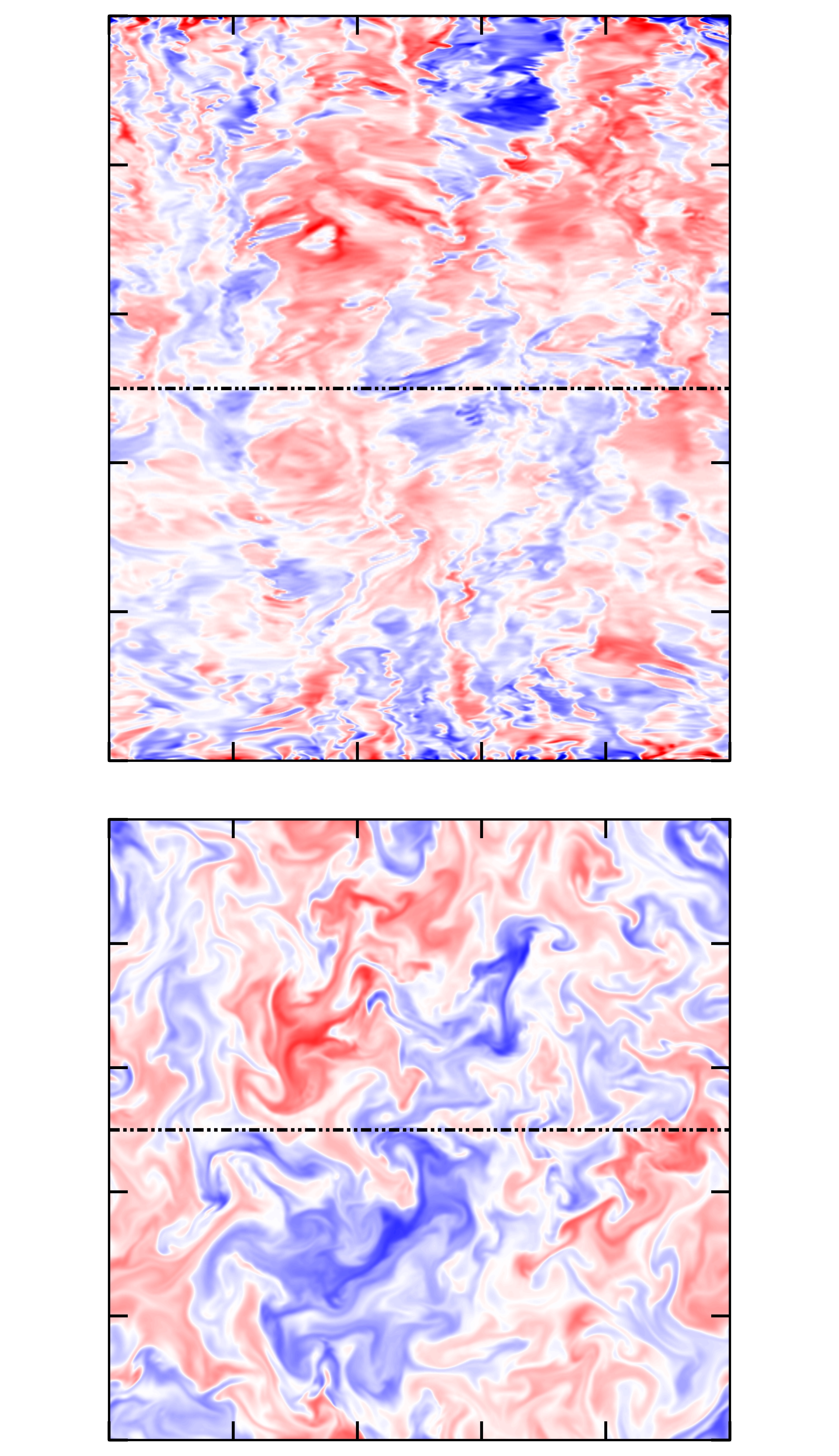}} \\
		%\vspace{-5mm}
		\subfloat[]{\includegraphics[width=0.33\textwidth]{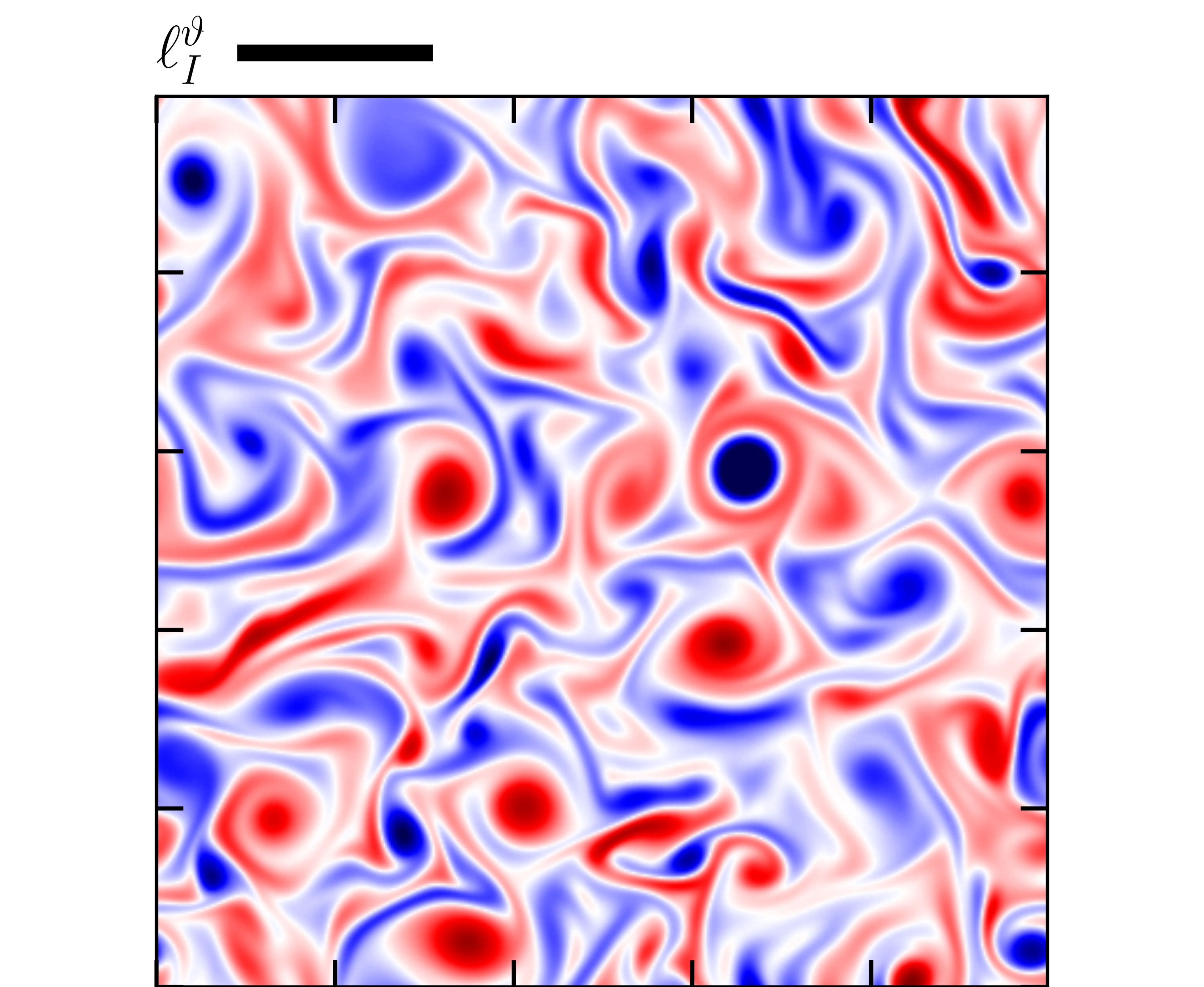}}
		%\subfloat[]{\includegraphics[width=0.53\textwidth]{R100temp_midplane_wscale.eps}} \\
		\subfloat[]{\includegraphics[width=0.33\textwidth]{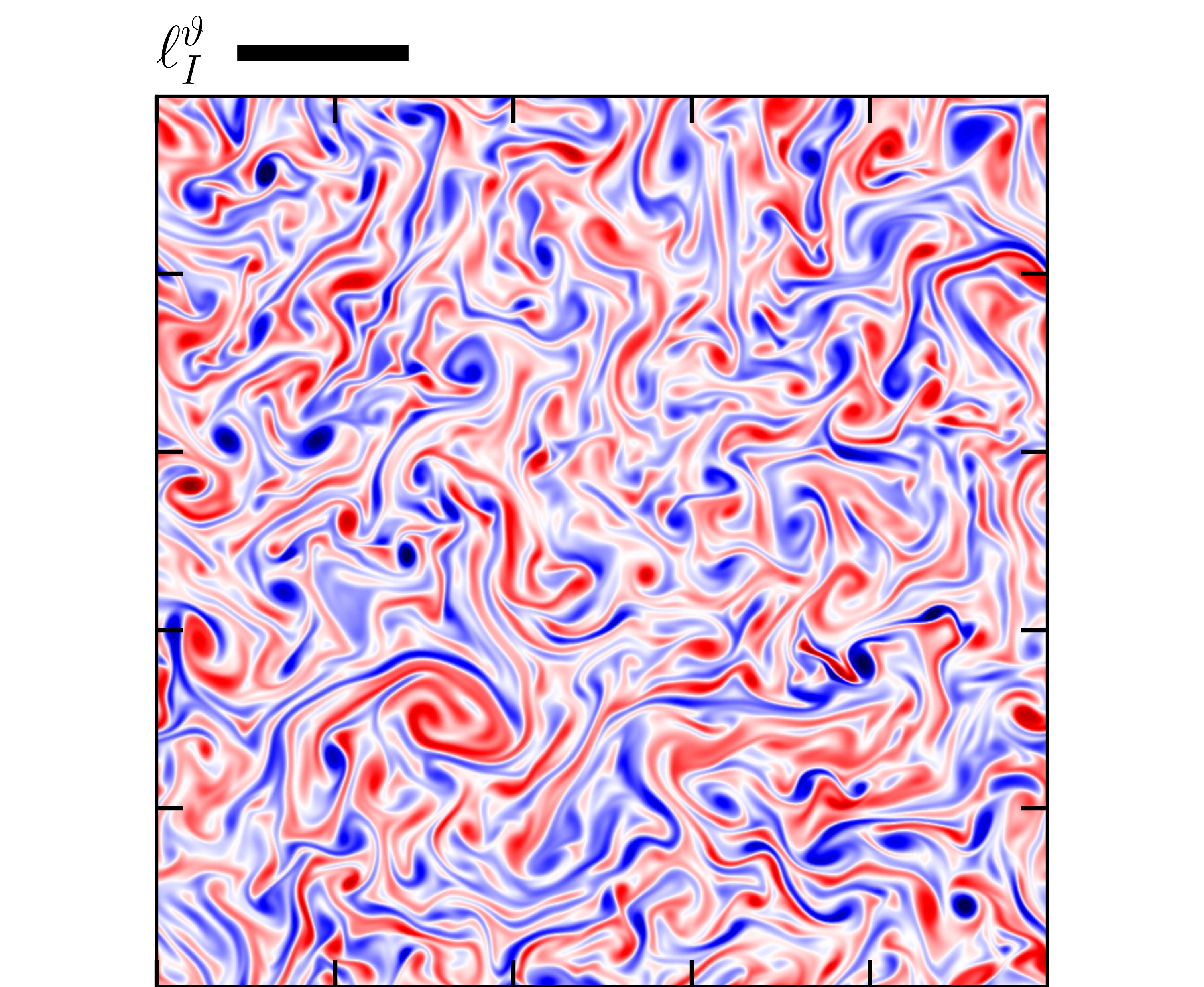}}
		\subfloat[]{\includegraphics[width=0.33\textwidth]{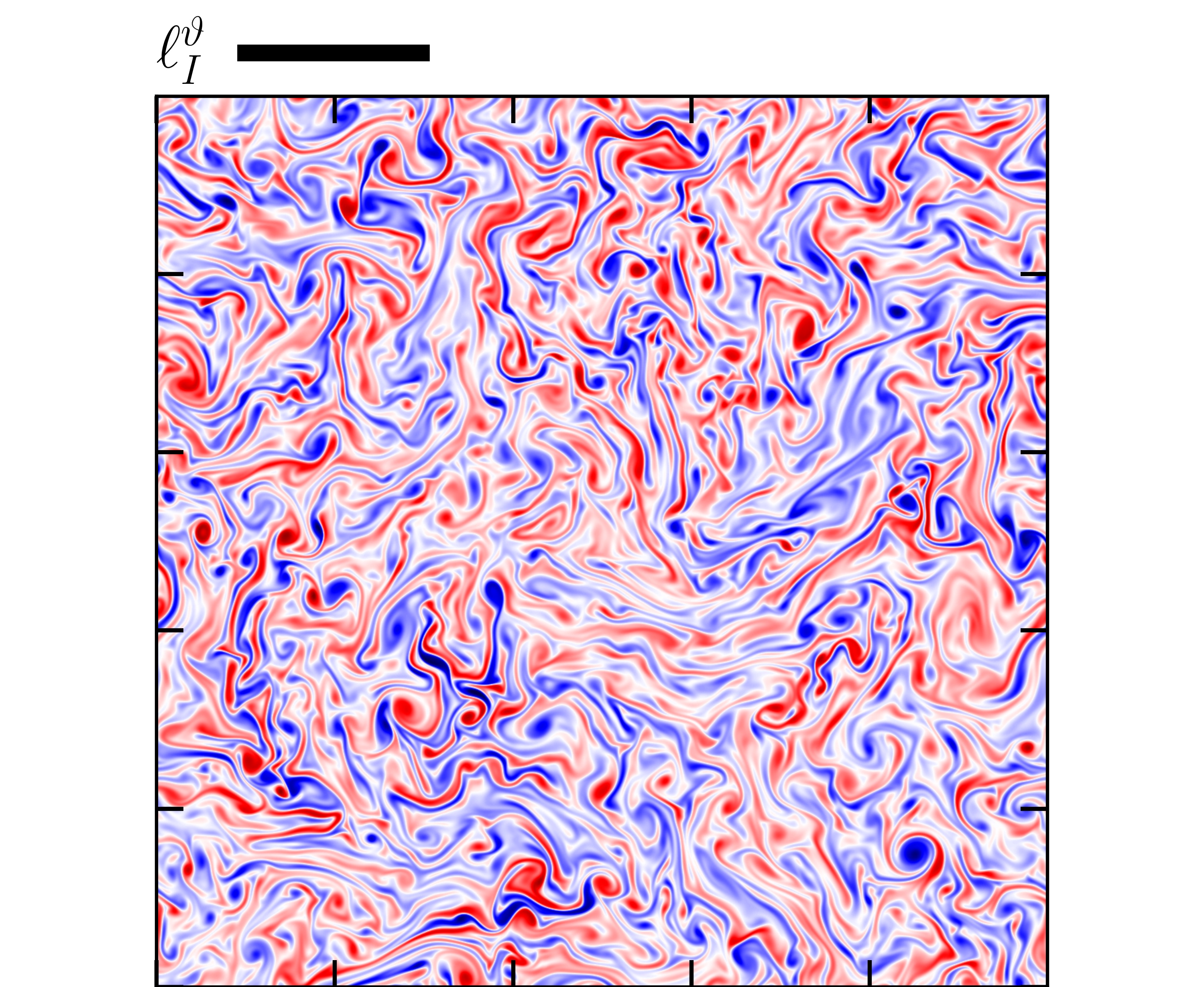}}		
	\end{center}
	\caption{
		Snapshots of the fluctuating temperature: (a,d) $\Rat=40$; (b,e) $\Rat=160$; (c,f) $\Rat=280$.
		Integral temperature scale is given by the black bars.
(a)-(c) Top row: $x$-$Z$ slices; bottom row: $x$-$y$ midplane slices. 
The units of $x$ and $y$ are given in number of critical wavelengths. Dot-dashed lines mark the intersection of the two planes ($x$-$Z$ and $x$-$y$). 
(d)-(f) Horizontal cross sections within the (upper) thermal boundary layer \textcolor{black}{(the corresponding depths are $Z = 0.985,0.9979,\text{ and } 0.9994$ for $\Rat = 40, 160, \text{ and }280 $ respectively).	}  }
\label{temp}
\end{figure}

\begin{figure}
	\begin{center}
		\subfloat[]{\includegraphics[width=0.33\textwidth]{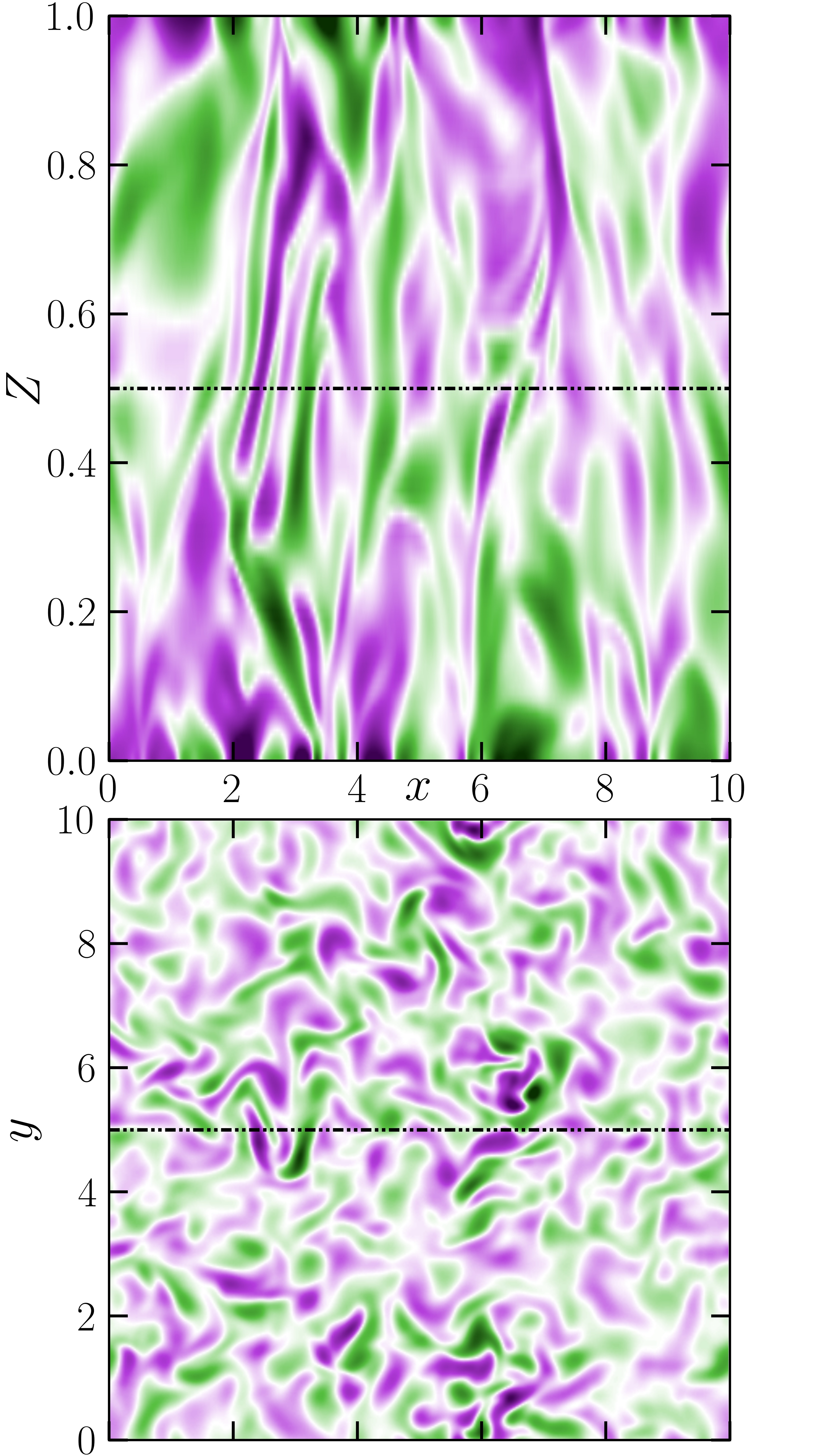}}
		%\subfloat[]{\includegraphics[width=0.53\textwidth]{R100temp_midplane_wscale.eps}} \\
		\subfloat[]{\includegraphics[width=0.33\textwidth]{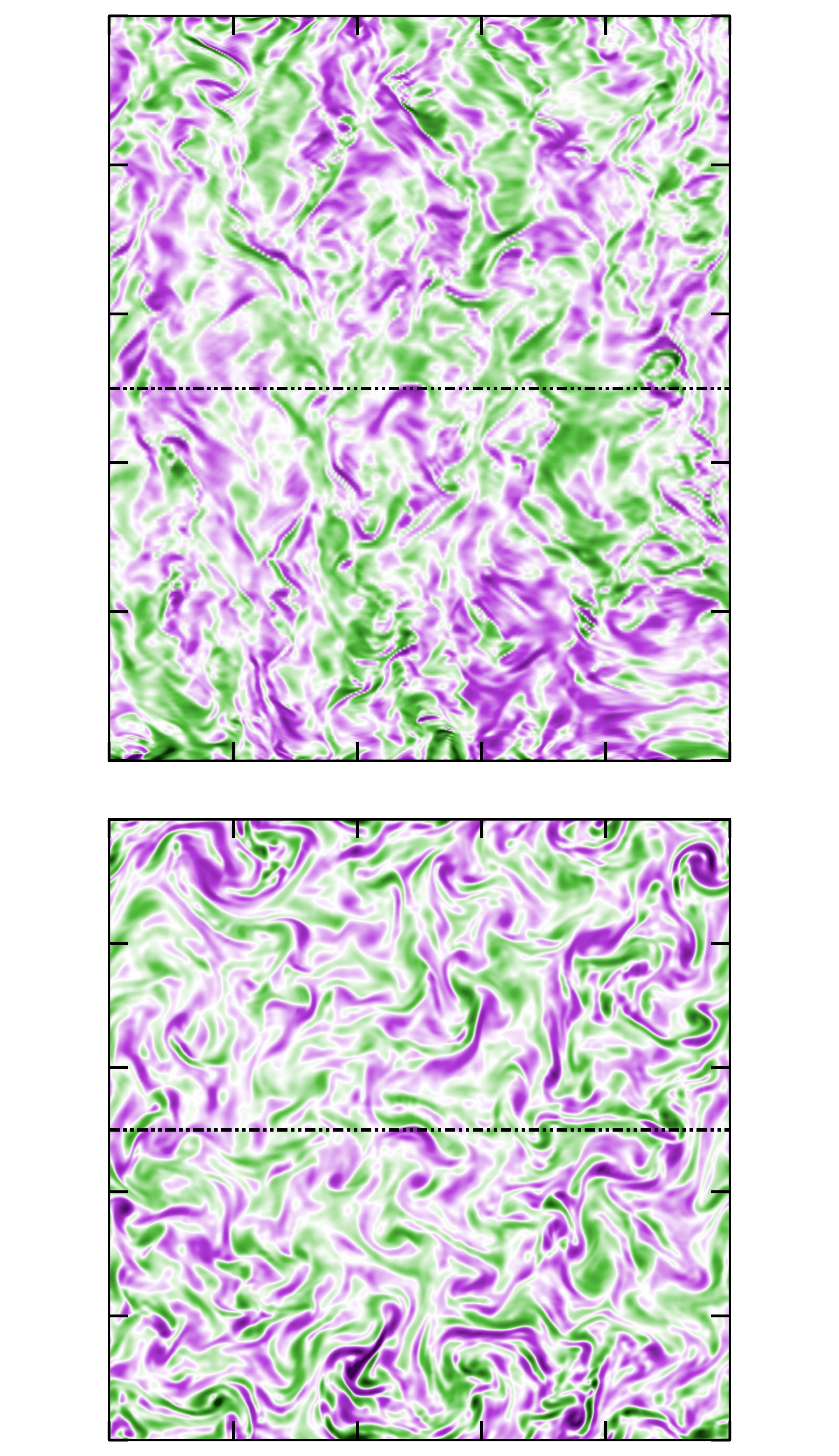}}
		\subfloat[]{\includegraphics[width=0.33\textwidth]{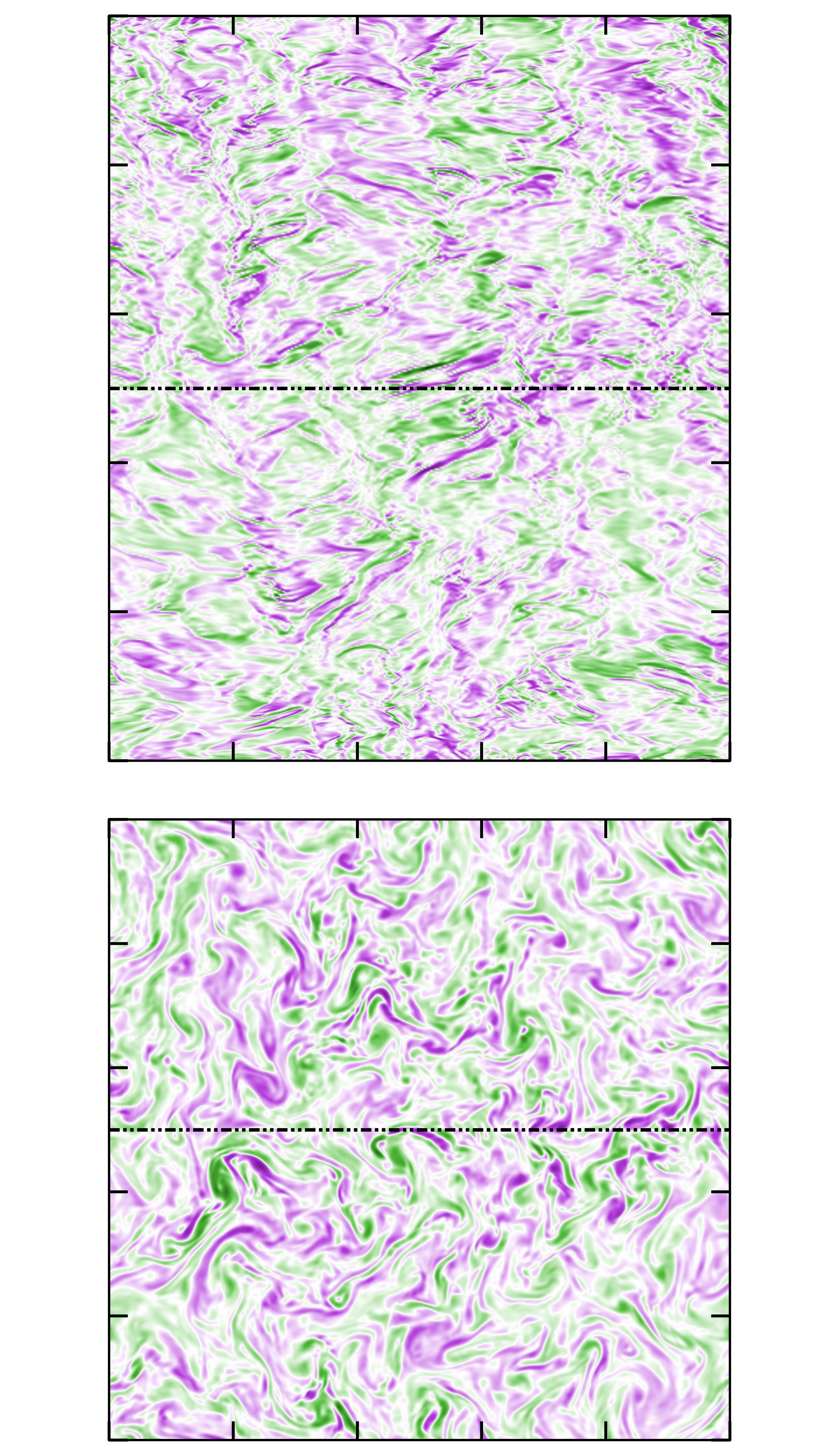}} \\
		%\vspace{-5mm}
		\subfloat[]{\includegraphics[width=0.33\textwidth]{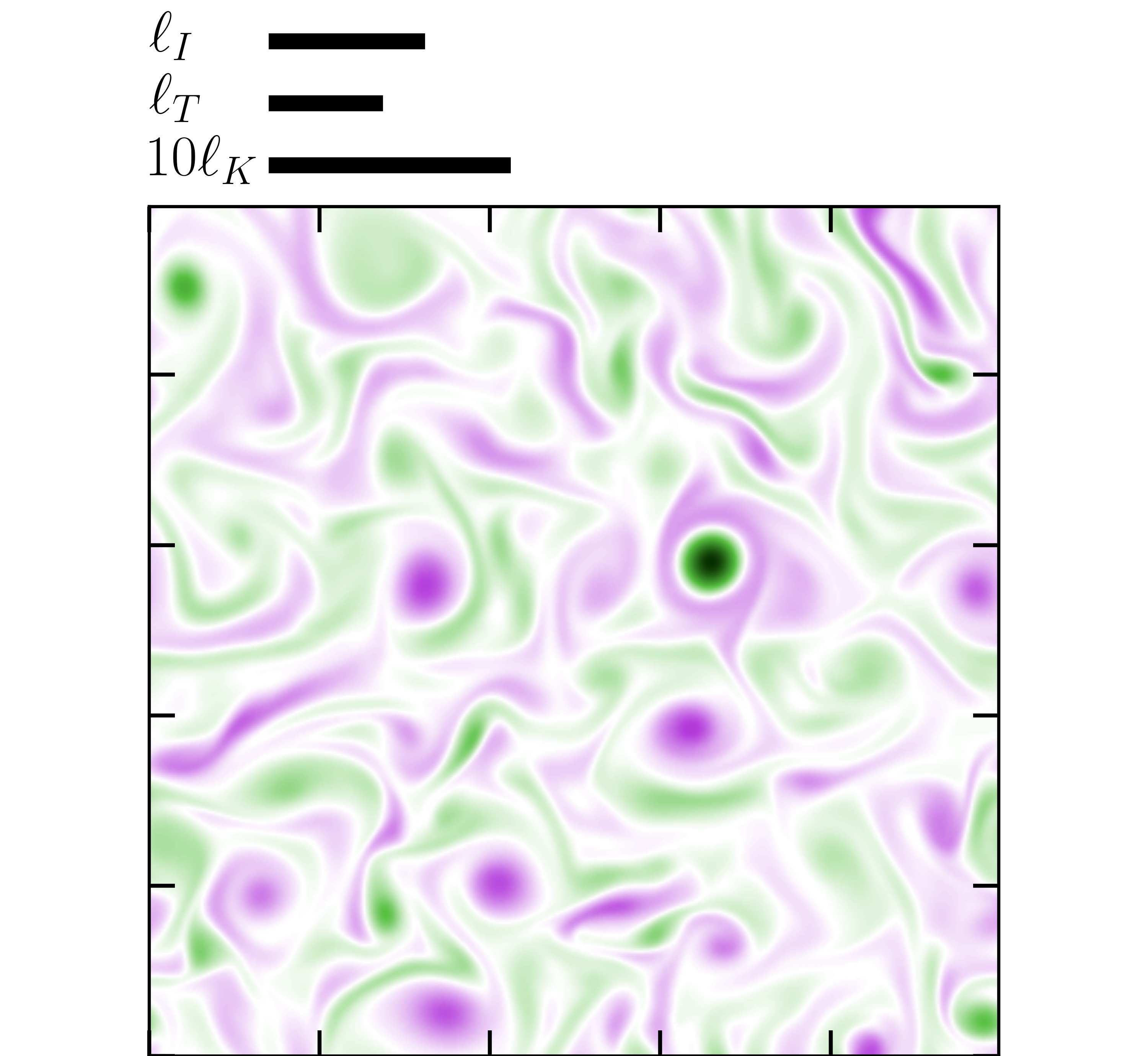}}
		%\subfloat[]{\includegraphics[width=0.53\textwidth]{R100temp_midplane_wscale.eps}} \\
		\subfloat[]{\includegraphics[width=0.33\textwidth]{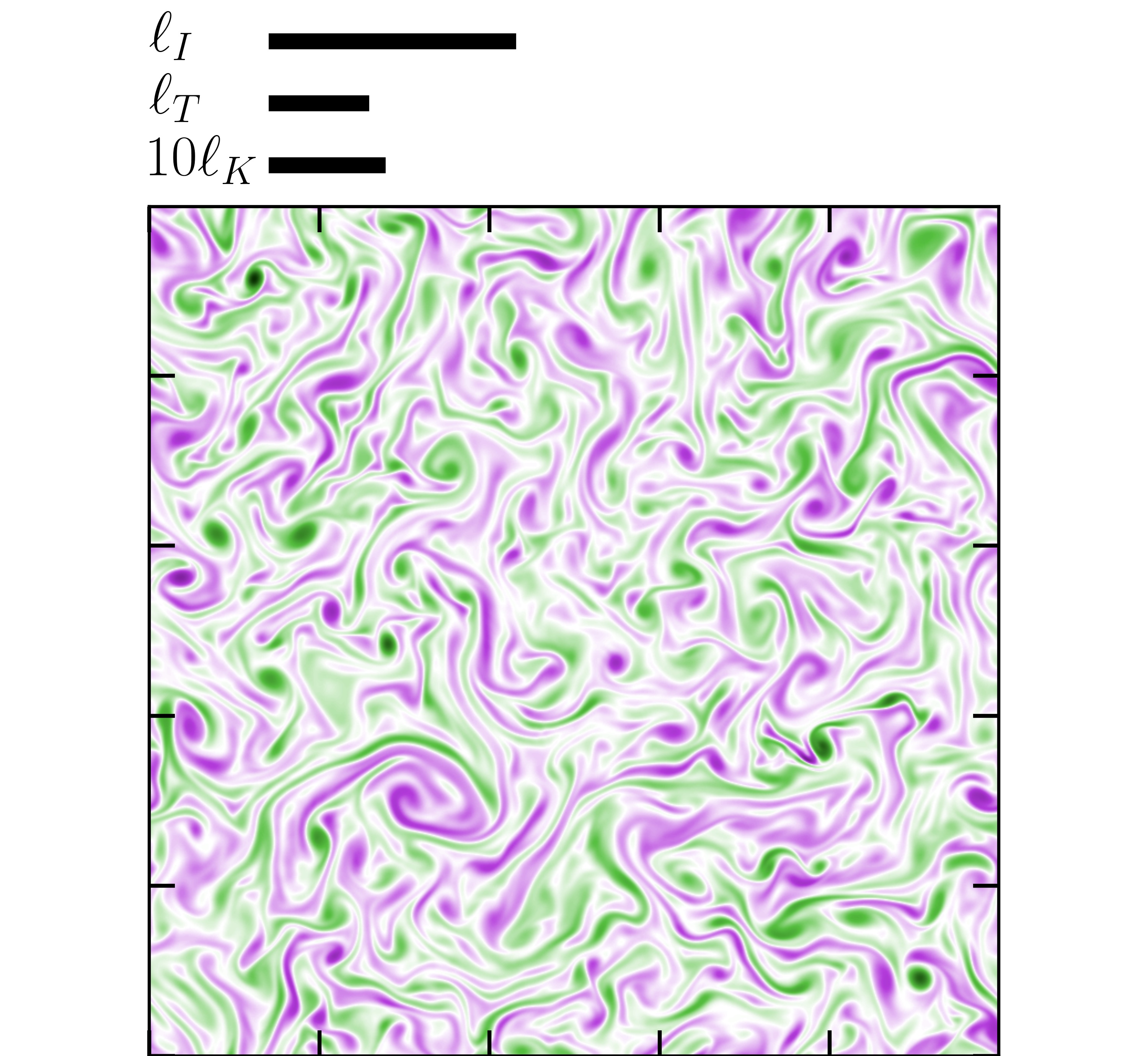}}
		\subfloat[]{\includegraphics[width=0.33\textwidth]{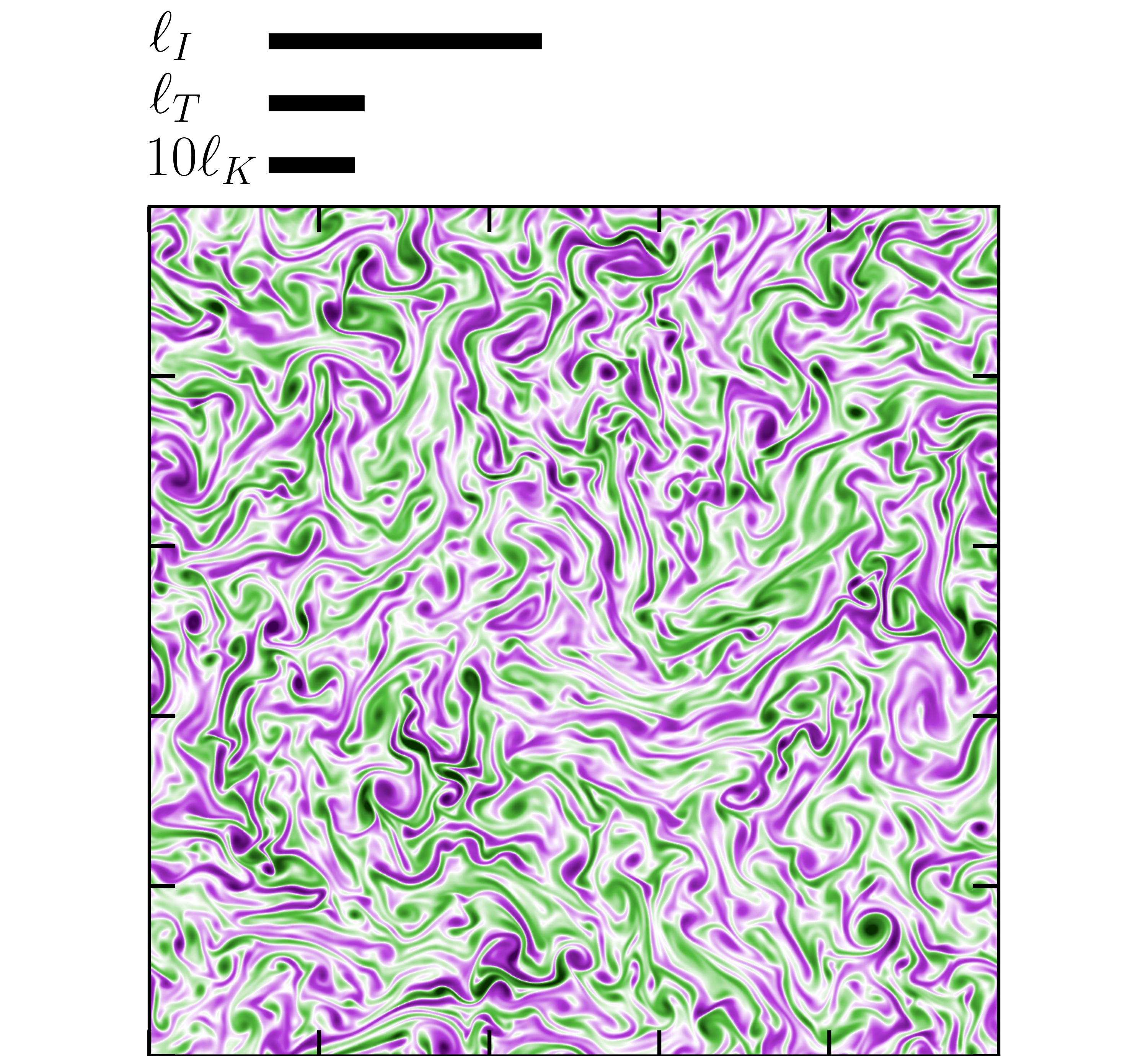}}		
	\end{center}
	\caption{
		Snapshots of the vorticity: (a,d) $\Rat=40$; (b,e) $\Rat=160$; (c,f) $\Rat=280$.
		Integral scale, Taylor microscale and ($10\times$) Kolmogorov scale are given by the black bars.
		Green and purple correspond to cyclonic and anti-cyclonic flows, respectively.	
	}
\label{vortz}
\end{figure}

\begin{figure}
	\begin{center}
		\subfloat[]{\includegraphics[width=0.33\textwidth]{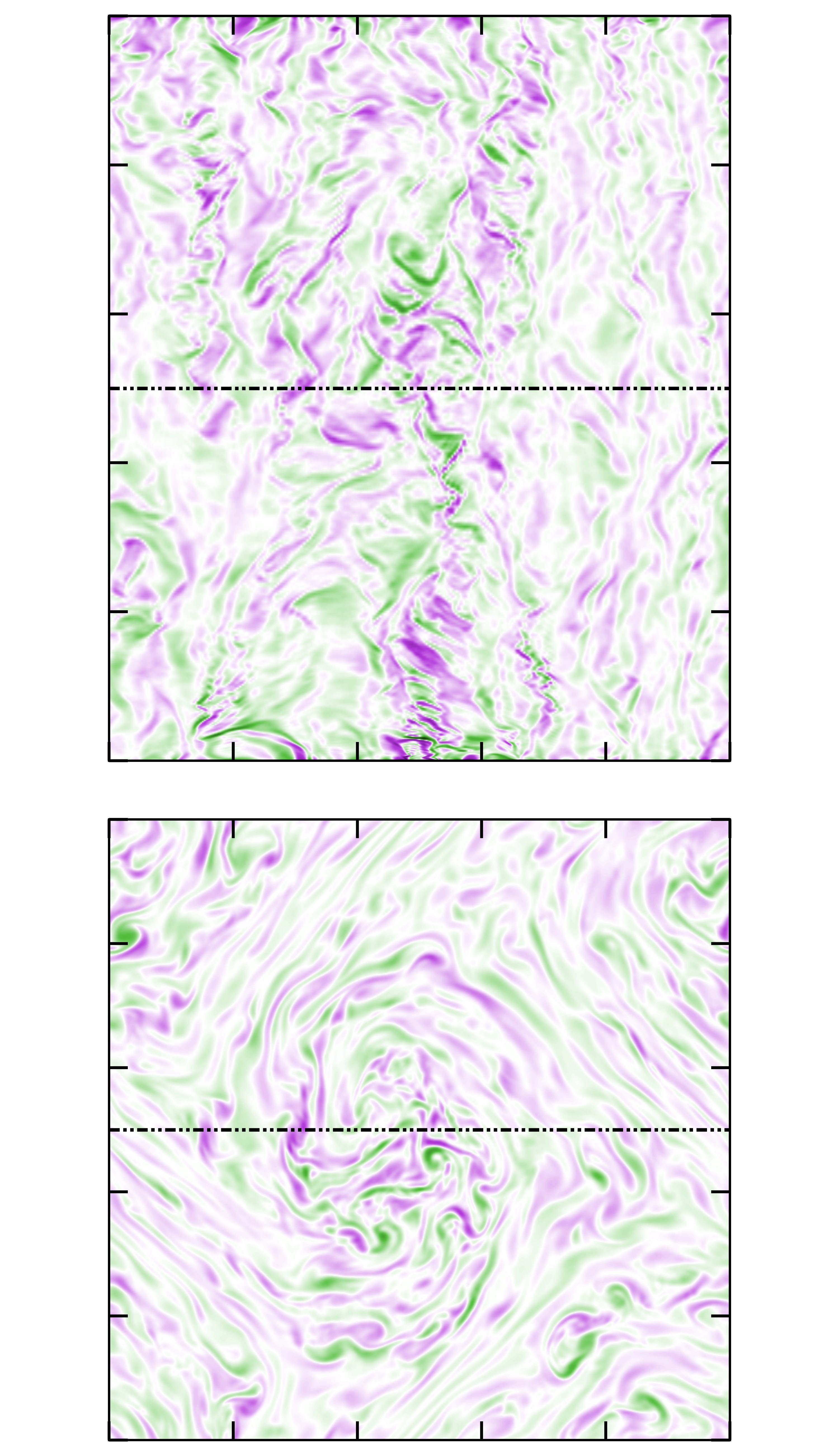}} \\
		\subfloat[]{\includegraphics[width=0.339\textwidth]{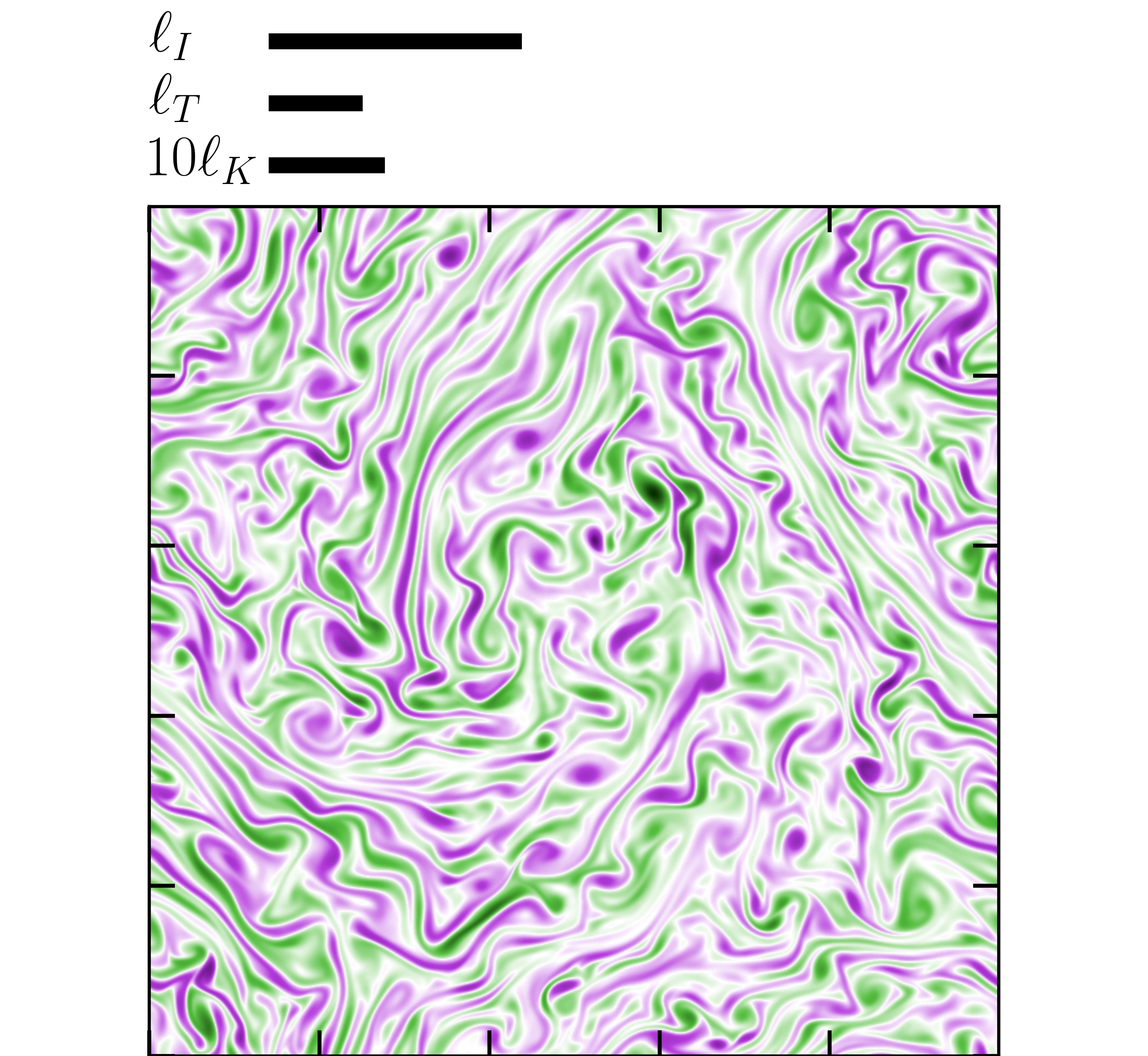}}
	\end{center}
	\caption{\textcolor{black}{Snapshots of the baroclinic vorticity from a case in which the depth-averaged flow is retained ($\gamma=0$, $\Rat = 160$). Note the presence of the dipolar large scale vortex (LSV) in the bottom of panel (a) and in panel (b).}
	}
\label{vortz_bc}
\end{figure}

Instantaneous physical space visualizations of the fluctuating temperature and vertical vorticity are shown in Figures \ref{temp} and \ref{vortz}, respectively, for $\Rat=40$ (a,d), $\Rat=160$ (b,e) and $\Rat=280$ (c,f). 
For each value of $\Rat$, all visualizations are taken at the same instant in time.
Vertical slices are shown in the top row of panels (a)-(c) and horizontal slices taken from the mid-plane of the fluid layer are shown in the bottom row of panels (a)-(c); the dot-dashed lines in (a)-(c) mark the intersection of the vertical and horizontal planes in the top and bottom rows, respectively.
%The top (bottom) row show the vertical (horizontal) locations of the slices in the bottom (top) row. 
Panels (d)-(f) show horizontal slices at depths in which $\vartheta_{rms}$ reaches its maximum value, i.e.~\textcolor{black}{at the edge of} the thermal boundary layer.
The visualizations confirm that the characteristic size of the large scale flow patterns \textcolor{black}{in the fluid interior}, as quantified by the integral length scale, is weakly dependent on the Rayleigh number \textcolor{black}{since all three cases show similar large scale structure}. 
%\textcolor{red}{KJ; I think is true for the interior but not the boundary layer. Not sure what features the read should be looking at. TO: seems like characteristic scale is weakly dependent on $\Rat$ for both to me}
Moreover, these large scale structures show significant axial coherence even in the absence of depth-averaged flows.
Even for $\Rat =280$, flow structures that span \textcolor{black}{nearly} the entire depth can be identified. However, it can be observed in both fields that the vertical length scale of coherence within the interior decreases with increasing $\Rat$. This is discussed in the context of non-local CIA theory in the final section.
The thermal boundary layer visualizations show evidence of strong spatial correlation between the thermal and vortical fields. While the size of the largest vortical structures remain largely insensitive to $\Rat$, it is observed that the filamentary structures become increasingly finer which is consistent with the theoretical findings in Ref.~\citep{kJ12b}. For the vortical field, this fine scale is imprinted into the interior whereas the interior spatial structure of the temperature field is observed to be de-correlated with the boundary. In the next section, we associate this effect to the thermal field behaving as an advective-diffusive scalar stirred by the vortical field and thermally dissipated. 

\textcolor{black}{Figures \ref{vortz_bc} (a,b) show the baroclinic vorticity for $\Rat = 160$ where the LSV is retained; these visualizations should be compared with the middle column in figure \ref{vortz} [i.e.~panels (b,e)]. The presence of the large scale vortex is apparent even in the baroclinic dynamics in both the midplane and the boundary layer, however the smaller length scales remain similar to those shown in figure \ref{vortz} (b,e).
}

%(a)-(c) where instantaneous renderings of the fluctuating temperature at the horizontal and vertical mid-planes ($y \approx 0.5$ and $Z\approx 0.5$) are plotted for three different values of $\Rat$. 
%We recall that the horizontal dimensions of the simulation domain is fixed at 10 critical wavelengths for all cases. 
%Figure \ref{temp} (d)-(f) shows the flow structure within the thermal boundary layer. 
%Here the finer structures for high $\Rat$ are more visible.

%These visualizations illustrate that while finer scale flow structures emerge with increasing $\Rat$, the largest structures remain approximately the same size. 

%If we interpret $\ell_{T}$ as the longest length scale at which viscosity can deform eddies, this relationship makes sense. 
%Eddies larger than $\ell_{T}$ would be inviscid and thus subject to the Taylor-Proudman constraint.
%At each timestep, these eddies are supressed along with the rest of the barotropic component. 
%\textcolor{red}{wrt. the vertical profiles, I don't really know what to say about the integral scale.}
%A possible interpretation of the slow increase of the integral scale with $\Rat$ is that it represents the increase in the propensity for linear regions (continuous filaments) of the flow structure. \textcolor{red}{Some of these statements are unclear/vague}

\subsection{Balances}

\begin{figure}
	\begin{center}
		\subfloat[]{\includegraphics[scale=0.3]{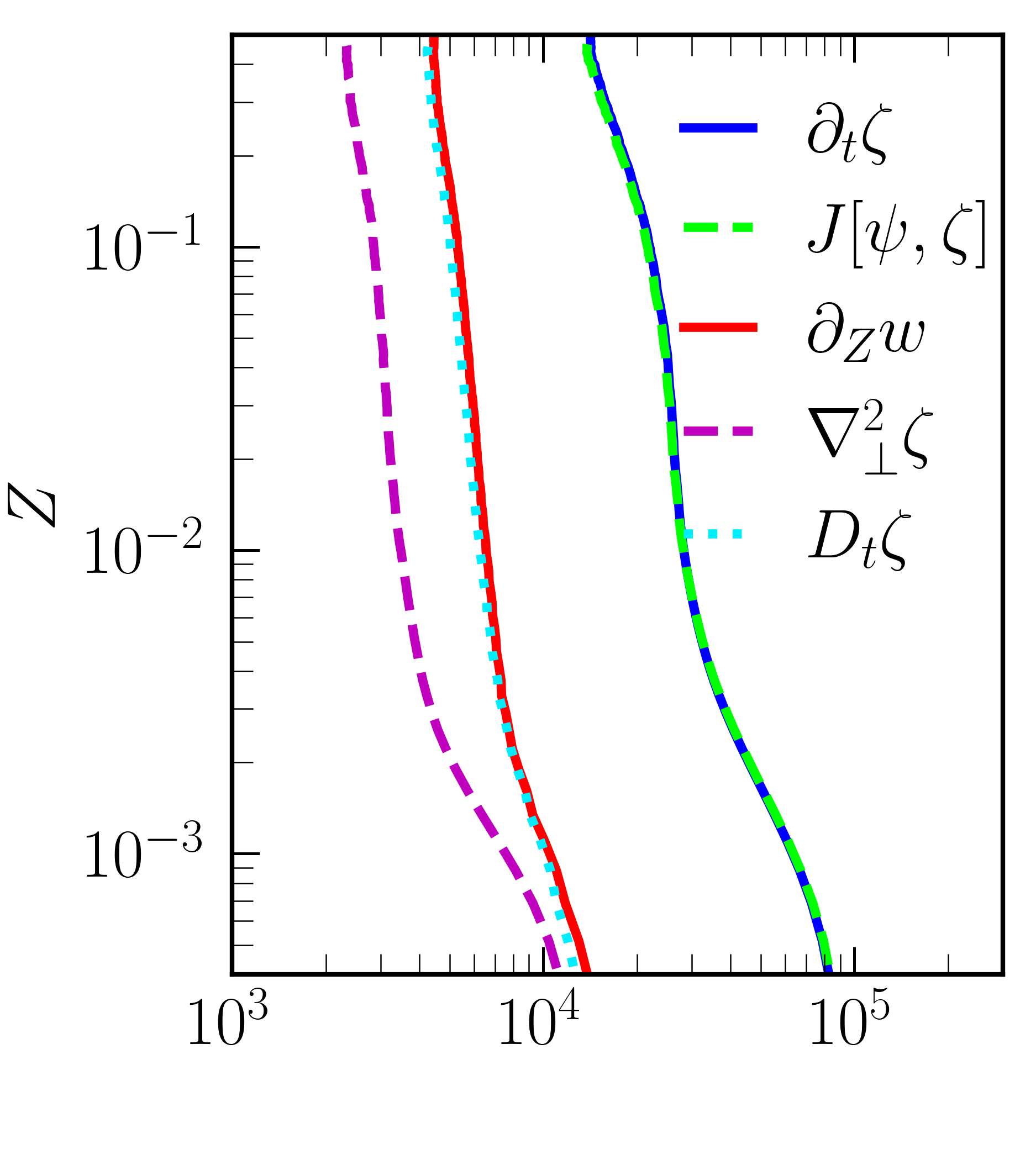}\label{sub:vortz}}
		\subfloat[]{\includegraphics[scale=0.3]{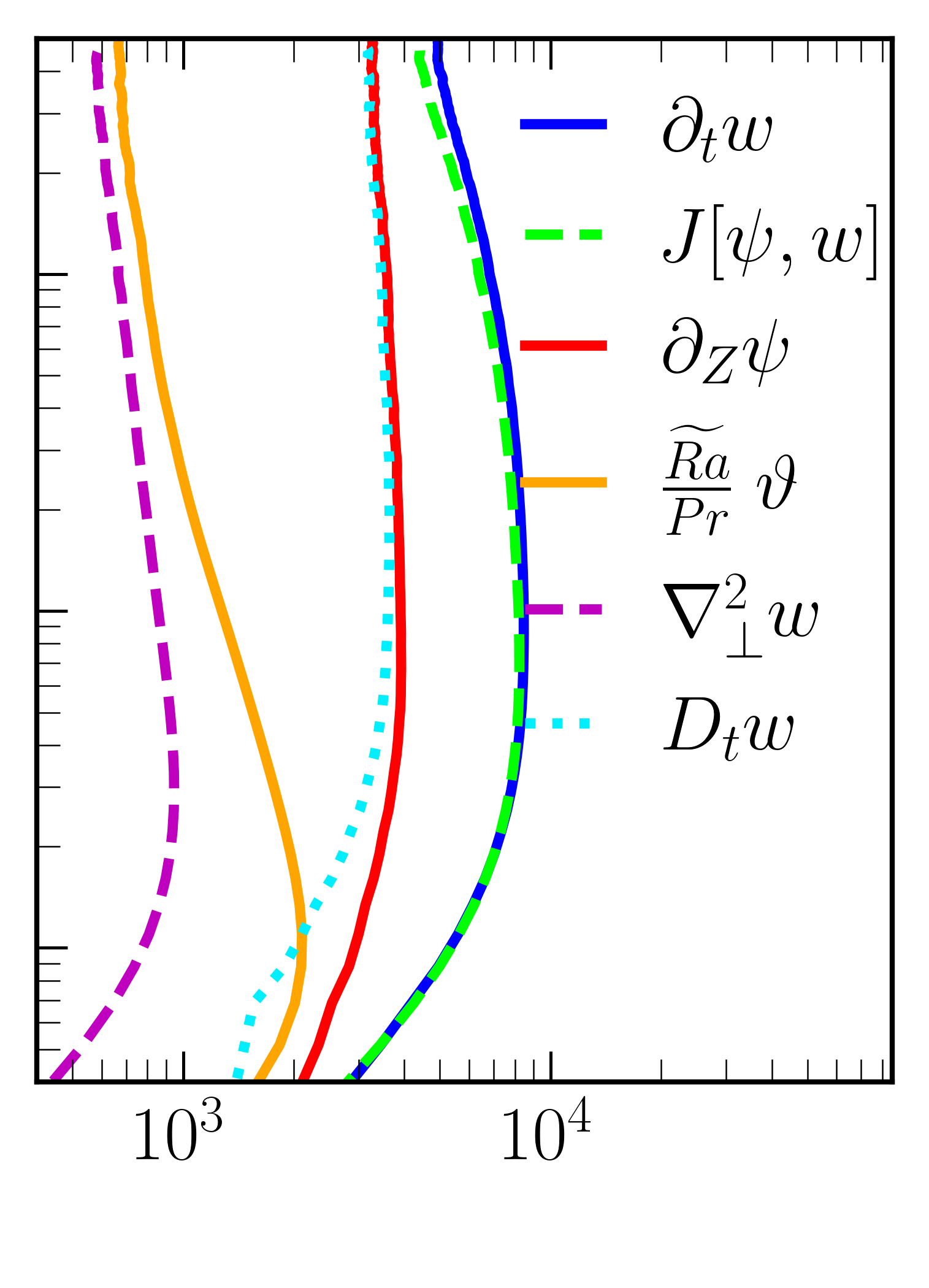}\label{sub:veloz}}
		\subfloat[]{\includegraphics[scale=0.3]{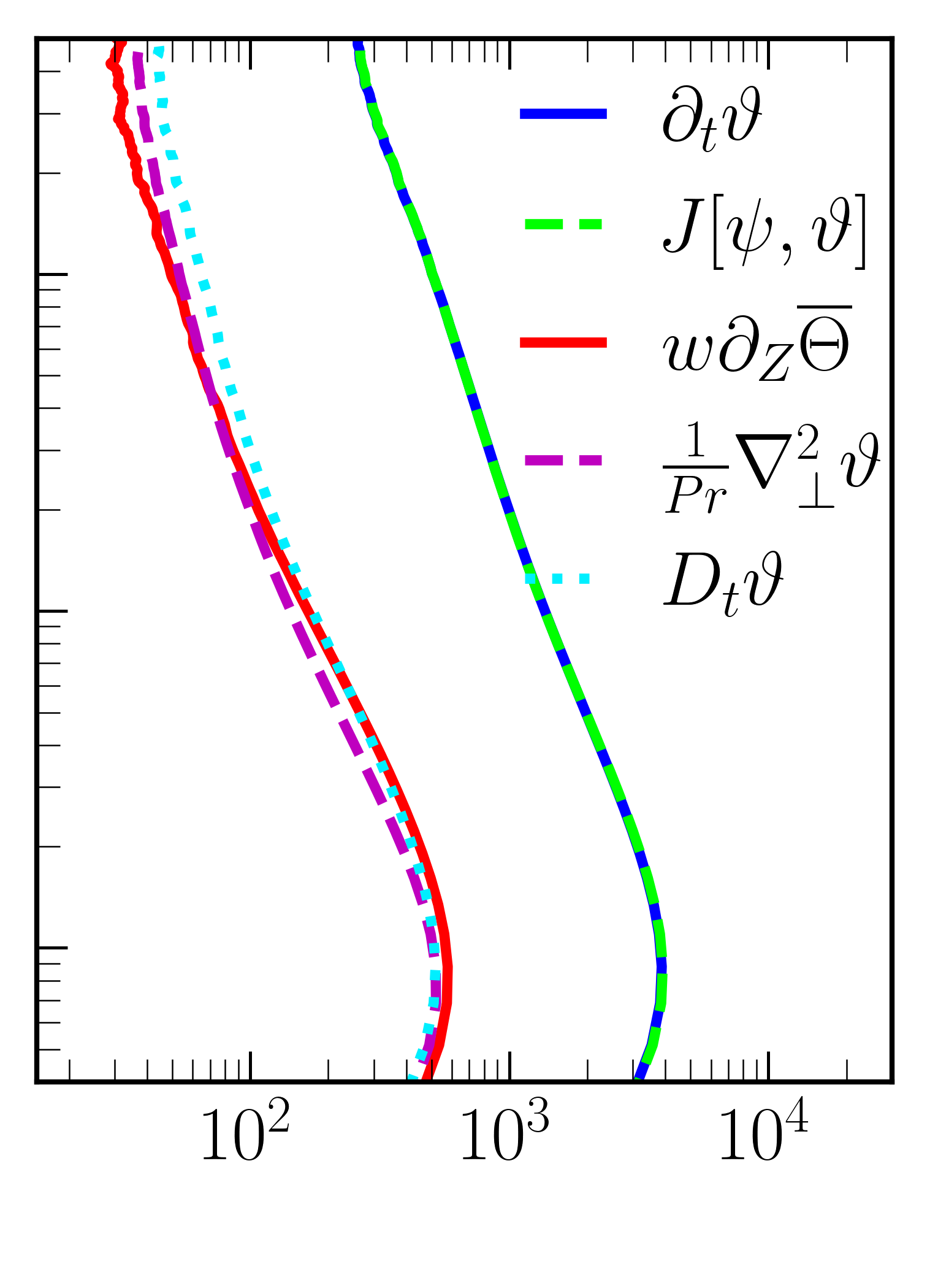}\label{sub:temp}} 
%		\subfloat[]{\includegraphics[scale=0.3]{R200Force_Balance_vortz_bl.png}\label{sub:vortz_bl}}
%		\subfloat[]{\includegraphics[scale=0.3]{R200Force_Balance_veloz_bl.png}\label{sub:veloz_bl}}
%		\subfloat[]{\includegraphics[scale=0.3]{R200Force_Balance_temp_bl.png}\label{sub:temp_bl}}
%		\hspace{100 mm}	
	\end{center}
	\caption{Logarithmic vertical profiles of rms values of each term in the governing equations for $\Rat = 200$: (a) vertical vorticity; (b) vertical momentum; (c) fluctuating temperature.}
	\label{F:Force_Balances}
\end{figure}

In an effort to understand the scaling behavior of the various quantities discussed thus far, we compute rms values of all terms appearing in equations \eqref{eqn:barotropicvort}-\eqref{eqn:barotropictemp}. Vertical profiles of these quantities are shown in Figure \ref{F:Force_Balances} for $\Rat = 200$, which is a value that is representative of the turbulent regime. The data shown in each plot is computed by squaring each term in the respective equations, averaging this quantity over the horizontal plane, taking the square root, and then averaging in time. The material derivative $D_{t}$ is the rms of the sum of all forcing terms with the advective terms subtracted off (\textit{i.e.} for the vorticity equation \eqref{eqn:barotropicvort}, rms($D_{t}\zeta$) = rms($\partial_{Z}w + \nabla_{\perp}^{2}\zeta$)). Note that a logarithmic scale is used on the vertical axis of each figure to illustrate the differences in balances that occur in the interior with those that occur within the thermal boundary layer. 
We find that advection and the time derivative terms are largest in each of the three equations throughout the fluid layer, which agrees with the results of Ref.~\cite{sM21} for simulations in which depth-invariant flows were present. 
However, the dynamics are controlled by the material derivative, rather than the time derivative and advection separately. 
Panel (a) shows that  $D_t \zeta$ and $\partial_z w$ are nearly identical in magnitude throughout the fluid layer. However, while smaller, the diffusion of vorticity remains comparable in magnitude to these two terms; in the fluid interior ($0.5 > Z > 0.1$) we find rms($\nabla_\perp^2 \zeta$) $\sim 2 \times 10^3$, whereas rms($D_t \zeta$) \textcolor{black}{$\sim 4 \times 10^3$} and rms($\partial_z w$) \textcolor{black}{$\sim 4 \times 10^3$}. Within the thermal boundary layer ($Z \lesssim 10^{-3}$) we find that all terms in the vorticity equation are important, indicating that the large amplitude vortical motions are directly influenced by viscous diffusion at the boundaries. 

For the vertical momentum equation balances shown in Figure \ref{F:Force_Balances}(b) we find that $D_t w$ and $\partial_Z \psi$ are of nearly identical magnitude in the fluid interior, whereas the buoyancy force and diffusion terms are comparable in magnitude and the smallest of all terms in the interior. These results show that the vertical pressure gradient acts as the dominant driver of vertical motion in the interior. In the thermal boundary layer we find that all terms in the vertical momentum equation become comparable in magnitude, though as a consequence of impenetrability, diffusion remains the smallest of all terms. Figure \ref{F:Force_Balances}(c) shows a tendency in the interior for the horizontal material advection of fluctuating temperature to be balanced by horizontal thermal diffusion, whereas all terms become comparable in magnitude within the thermal boundary layer.

\begin{figure}
	\begin{center}
        \includegraphics[scale=0.4]{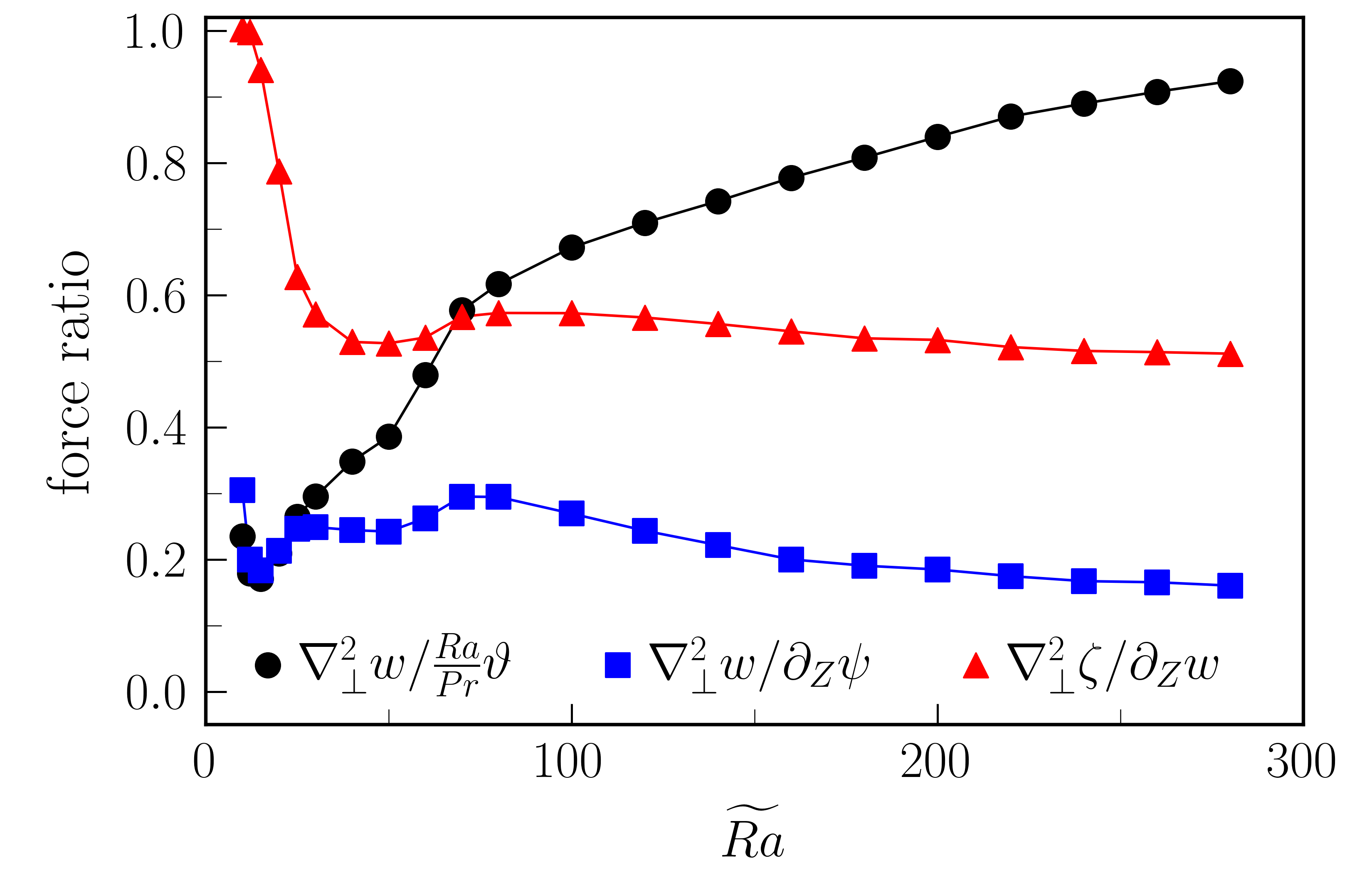}\label{sub:ratio}
	\end{center}
	\caption{Select ratios of rms forces from vertical momentum and vorticity equations. In the vertical momentum equation, the ratio of diffusion to buoyancy approaches unity. The ratio of diffusion to \textcolor{black}{the} pressure gradient (vortex stretching) in the vertical momentum (vorticity) equation is a slowly decreasing function of $\Rat$ for large $\Rat$. 
	}
	\label{F:Force_Ratio}
\end{figure}

%\textcolor{black}{UPDATE THIS PLOT TO GET ALL SELECTED RATIOS}.
Figure \ref{F:Force_Ratio} show ratios of rms values of various terms in the vertical momentum and vertical vorticity equations as function of $\Rat$. 
The rms values are the depth-averaged values of vertical profiles similar to those shown in Figure \ref{F:Force_Balances}(b). 
We find that the ratio of the vertical viscous force to the buoyancy force approaches unity as $\Rat$ is increased. Given the scaling behavior of the rms fluctuating temperature, $\vartheta_{rms}$ (as shown in Figure \ref{F:temp}(a)) this data indicates that the viscous force increases at a rate faster than $\Rat^{1/2}$ until unity is reached in the force balance. 
This balance between the viscous force and the buoyancy force is attributed to the observation that the fluid interior controls heat transport in the large Rayleigh number regime of rapidly rotating convection \citep{kJ12b}. 
The ratios of the viscous force to the pressure gradient force and the diffusion of vertical vorticity to vortex stretching are slowly decreasing functions of $\Rat$. From our data, while it appears that the magnitude of the globally averaged influence of viscosity is subdominant from that of the pressure gradient or vortex stretching in the limit of large $\Rat$, it is unclear whether asymptotic subdominance holds for $\Rat\rightarrow\infty$.

\section{Discussion and Conclusions}
\label{S:conclusions}

The QG equations  \eqref{eqn:barotropicvort}--\eqref{eqn:barotropicdff} represent the asymptotic low Rossby number limit of the buoyantly forced Navier-Stokes equations. 
Simulations of these equations were performed in which the depth-invariant flows 
%\textcolor{red}{including the large scale vortices} 
were entirely suppressed. 
The suppression was done to isolate the dynamics of the small scale convection and to enable simulations at previously inaccessible values of $\Rat$. A comparison was made with data from previously published simulations of the asymptotic model in which large amplitude, depth-invariant flows (e.g.~LSVs) were present. 
This comparison has allowed for additional insight into the physics of small scale, QG convective turbulence, which drives the inverse kinetic energy cascade in this system. Asymptotically reduced Rayleigh and Reynolds numbers up to $\Rat = 280$ and $\Ret \approx 100$ were simulated, which represents the most extreme parameter regime accessed to date for QG convection in either the plane layer or spherical geometry. 
%For comparison, at a representative value of $Ek\sim 10^{-15}$ for the Earth's outer core, this equates to $Ra\sim10^{22}$ and $Re\sim 10^7$.
Recent DNS studies have reached small scale Reynolds numbers up to $\Ret \approx 33$ in the plane layer \citep{mY22a,mY22}, and up to $\Ret \approx 24$ in a spherical shell geometry \citep{nS17}. Ref.~\cite{cG19} used a non-asymptotic two-dimensional model in the equatorial region of a full sphere to reach $\Ret \approx 20$. For context, estimates suggest that characteristic flow speeds in the Earth's outer core yield reduced Reynolds numbers of $\Ret \approx 10^3-10^4$, clearly indicating that it is necessary to understand the physics of QG convection at large values of the reduced parameters. 

A subset of simulations were performed in which the horizontal dimensions of the simulation domain were systematically varied.
In agreement with previous studies of non-rotating convection \citep[e.g.][]{rS18}, we find that \textcolor{black}{globally}
%different quantities show a different dependence on the simulation domain size (Figure~\ref{figure:box_dim}). 
averaged statistics, such as the Nusselt number and the Reynolds number, converge rapidly for horizontal dimensions that are larger than ten critical horizontal wavelengths (Figure~\ref{figure:box_dim}). 
The \textcolor{black}{calculated length scales remained} approximately constant across the different domain sizes. 
%The integral length scale of the flow field converges to a constant value with increasing domain size, but at a rate that may be influenced by the particular value of $\Rat$. However, the relative differences observed are small for our investigated range of $\Rat$; a conclusion that is supported by the rapid convergence of the Reynolds number with increasing domain size.  
%\textcolor{red}{Perhaps a discussion of the box size and relationship to sphere would be useful. For the spherical geometry, reducing the Ekman number becomes equivalent to increasing the box size in some sense. TO, by this do you mean that it makes the critical wavelength smaller wrt box dimension? I'm not sure how constructive this comparison will be, as it might give the impression that changing box size actually changed your parameters. Maybe we just make a comment about the size of the domain that we would be simulating for the Earth ($10\times D \times E^{1/3} \sim 200 m?$).}

Our investigation finds evidence that in the absence of depth-invariant flows both the heat and momentum transport approach asymptotic scaling regimes for large Rayleigh numbers (Figure~\ref{F:Nu_Re}), \textcolor{black}{though these scaling regimes occur only at the largest accessible values of $\Rat$.} When generalized to non-unity Prandtl numbers, \textcolor{black}{these scalings are} $Nu \sim Pr^{-1/2}\Rat^{3/2}$ and $\Ret \sim \Rat/Pr$. 
When recast in terms of large scale quantities these scalings become $Nu \sim Pr^{-1/2} Ra^{3/2} E^2$ and $Re \sim Ra E/Pr$, which represent diffusion-free scalings when interpreted on the large domain scale \citep{jmA20}.
%\textcolor{black}{For $Nu$, the data suggests this asymptotic scaling, but it is possible that a large enough $\Rat$ regime has not been reached. 
%	The momentum transport data is less convincing. For $\Rat>10,$ a power scaling of $\Rat^{1.33}$ was found to fit the data.  
%Renormalizing  $\Ret$ by $\Rat$ as suggested by CIA theory appears to only flatten the data for $\Rat\geq 220.$ Further simulations would be required to confirm whether this scaling continues to hold. 
%}

% are consistent with a diffusion-free scaling resulting from a Coriolis-Inertia-Archimedean (CIA) balance \citep{cG19,jmA20}. 
The simulations that include depth-invariant flows exhibit heat and momentum transport that is more efficient than the aforementioned diffusion-free scalings \citep{sM21}. This diffusion dependence is of interest when considering  applications to geophysical and astrophysical systems in which large-scale flows, such as zonal jets and vortices, are present. 
%The findings from the present small-scale QG convection study therefore provides an important benchmark for comparison.

%
% MAC: not totally convinced the section below is necessary, or it needs more explanation
%
% The Reynolds number scaling indicates that fluid flow velocities achieve rotational free-fall $U_{r\!f\!f}=g\alpha\Delta T/2\Omega$, whereas the Nusselt number scaling indicates that non-dimensional temperature fluctuations obey $\theta \equiv E^{-1/3} (\theta^*/\Delta T) \sim  \lb\Rat/Pr\rb^{1/2}$. \textcolor{red}{MAC: strange to switch back and forth between dimensional and non-dimensional}
%These scalings are also consistent with the large and small scale Reynolds number estimates
%\be
% Re = \left [ \frac{Ra}{Pr}\frac{Nu-1}{Pr}  \right ]^{2/5} E^{1/5}
%\quad \Longleftrightarrow \quad
%  \Ret = \left [ \frac{\Rat}{Pr}\frac{Nu-1}{Pr}  \right ]^{2/5} 
%\ee
%deduced from the exact relation \eqref{eq:edisspn} for the volume-averaged energy dissipation $\epsilon_u$ upon estimating it dimensionally as $\epsilon^*_u =  (\nu^3/\ell_\nu)\epsilon_u \sim u^{*3}/{\ell^*_I}$ (see also \citep{ek13b}). \textcolor{red}{This assumes that the large length scales control dissipation?}

Kinetic energy spectra and temperature variance spectra were computed for all values of $\Rat$ (Figure~\ref{F:spectra}). 
\textcolor{black}{
We find evidence of a Kolmogorov-like subrange in which the kinetic energy spectra scales with the horizontal wavenumber as $k^{-5/3}.$ }
%We do not find evidence of a trend toward a Kolmogorov-like inertial subrange in which the kinetic energy spectra scales with the horizontal wavenumber as $k^{-5/3}$; a stronger dependence closer to $k^{-5/2}$ is observed. 
%\textcolor{red}{KJ: again we should have a discussion of the inertial subrange}
\textcolor{black}{ However, since we find that the inertial scale is a viscous scale, we believe that this subrange is distinct from the classical picture of a Kolmogorov inertial subrange.}
The kinetic energy spectra show that there is a build up of energy at smaller wavenumbers as $\Rat$ is increased, though the energy contained in these larger scale structures remains comparable to that contained in scales similar to the critical wavelength. The temperature variance spectra exhibit similar behavior, indicating that the system remains forced on a length scale that is comparable to the critical wavelength at the highest Rayleigh number investigated.

The evolution of various length scales in the system was investigated for varying $\Rat$. 
We find that the integral length scale $\ell_I$ increases with $\Rat$, but at a rate that is slower than the diffusion free scaling $\propto (\Rat/Pr)^{1/2}$ suggested by a CIA balance. %\textcolor{black}{The integral scale reaches a local maximum at $\Rat =50$ before decreasing slightly up to $\Rat=70$, followed by a   before a short region of decrease and the slow increase.}
The Taylor microscale remains nearly constant with increasing $\Rat$, whereas the Kolmogorov length scale decreases with increasing $\Rat$.
\textcolor{black}{The temperature integral scale $\ist$ was was found to behave similarly to the integral scale, albeit with a more pronounced peak at $\Rat = 50$. Both integral length scales show increases for $\Rat>100$.}
One of the key findings of this investigation is that all of these length scales remain comparable to the linearly unstable critical wavelength that emerges at the onset of convection. 
We note that the scaling behavior of the integral scale and Taylor microscale are consistent with recent laboratory experiments of rotating convection \citep{mM21} and a numerical study of rotating convection-driven dynamos in the plane layer \citep{mY22}. 
These findings provide evidence that the broadening of the length scales in rotating convective turbulence is an extremely slow function of the Rayleigh number, and occurs in a fundamentally different manner in comparison to non-rotating turbulence. 
In particular, studies of non-rotating convection find that both the integral length scale and the Taylor microscale decrease rapidly with increasing Rayleigh number \citep[e.g.][]{mY21}.

An explanation for the aforementioned contrast to the non-rotating case resides in the fact that viscosity is a required ingredient in destabilizing a rotating fluid subject to an adverse temperature gradient in the presence of the Taylor-Proudman constraint  \citep{sC61}. 
\textcolor{black}{Even in the turbulent regime, the Taylor-Proudman constraint is relaxed on the length scale associated with unit $Ro$. In rapid rotation, this is a viscous length which scales as $ Ek^{1/3} $.}
Convective motions are therefore inextricably influenced by viscosity and, by definition, the Taylor microscale is tied to the linear instability scale $\ell^*_\nu= E^{1/3}H$. 
\textcolor{black}{We propose an alternate scaling motivated by linear theory.
For a fixed mean temperature gradient, we find that the most unstable mode grows with $\Rat$ like $k^{-8},$ suggesting a length scale of $\ell\sim\Rat^{1/8}.$
Our data (Figure~\ref{F:LS}\subref{fastest_mode}) for the integral scales at large $\Rat$ seem to agree with this scaling better than the dissipation-free scale predicted by CIA theory  $\propto \Rat^{1/2}.$}
This connection to linear theory is demonstrated by comparisons between the computed lengths and the marginal stability curve (Figure~\ref{F:LS}\subref{fastest_mode}). 
%It is found that the dependence of the integral length scale with increasing $\Rat$ follows the fastest growing, viscosity dependent, linear mode $\propto \Rat^{1/8}$, rather than dissipation-free scale predicted by CIA theory $\propto \Rat^{1/2}$. 
We note that the latter scaling resides at the long wavelength marginal stability boundary for convective onset. The comparisons with linear theory in which a mean temperature profile with gradient $\partial_Z \overline{\Theta} = -1$ might remain pertinent in the nonlinear regime given that the observed nonlinear profile saturates with a gradient $\sim -0.5$ (Figure~\ref{F:temp}\subref{dz_temp}). The permissible convectively unstable length scales within the bulk thus adhere to the same analytic structure as deduced from the marginal stability at linear onset.

Congruent with the lack of agreement between the measured integral length scale and that predicted by CIA theory, the force balances also do not show evidence of a CIA balance within the fluid interior, as might be predicted from scaling arguments. Instead we find a trend indicating that this balance is never approached as the system becomes more strongly forced since there is a subdominant balance between the buoyancy force and viscous diffusion of vertical momentum in the interior. The ratio of the rms of these two forces approaches unity as $\Rat$ is increased, which we attribute to the fact that the fluid interior controls heat transport in rapidly rotating convection \citep[e.g.][]{kJ12b}. The buoyancy force is strongest within the thermal boundary layer and comparable to all other forces, including viscous forces, in that region. Moreover, the strong, horizontal, vortical flows that are present at the boundary are balanced entirely by viscosity. Convective overturning motions in the interior are driven primarily by vertical pressure gradients, and vortical motions are driven by vortex-stretching. However, neither of these forces are sources of potential energy injection which must then be provided from within the boundary regions.

	\textcolor{black}{
	A possibility for the departure from a CIA balance is that there is no buoyancy term in the vorticity equation \eqref{eqn:barotropicvort} because gravity and rotation are antiparallel in this geometry. However, we might still expect a balance between buoyancy and the material derivative in the vertical momentum equation, which we do not observe. Nevertheless, it would be of interest to determine if the angle between the gravity and rotation vectors plays a role in the scaling behavior and balances that are observed in simulations. Such an investigation would be important for relating the results of plane layer and spherical simulations \citep[e.g.][]{tG23}. }
	
%	conducting simulations in which the gravity and rotation vectors are not aligned is a necessary task to determine if scaling behavior is dependent on
%	Furthermore, Ref. \citep{jA20} demonstrates that only a balance between the Coriolis acceleration and inertia in the vertical vorticity equation is necessary to show that $\ell_{I}\sim \vartheta.$ }
%\textcolor{red}{KJ: This is a confusing sentence, CI balance gives $\ell_{I}\sim Ro$. How does $\vartheta$ enter? TO: \citep{jA20}shows that if you just balance CI as well as the fluctuating temp gradient with the mean temp gradient that you can get $\ell_{I}\sim\vartheta.$ Am I confused here? Mike could you check}
%\textcolor{black}{We observe a $\Rat^{1/2}$ scaling for $\vartheta$ but not for $\ell_{I}.$ }

%\textcolor{red}{Maybe we add the discussion relating the linear behavior here.}

%\textcolor{red}{Add paragraph discussing comparison with linear stability behavior?}

\textcolor{black}{If we accept that the $Nu$ and $\Ret$ scalings are trending towards the dissipation-free scaling provided by CIA theory, then there appears to be a juxtaposition between global transport laws and the observed integral length scaling. This apparent contradiction} can be resolved by analyzing force balances within the fluid interior and boundary layers. 
Indeed, balances in the governing equations can often be used to derive various scaling laws for the Rayleigh number dependence of flow length scales. 
Towards this end we have computed the magnitudes of the various terms appearing in the governing equations of the asymptotic model (Figure~\ref{F:Force_Balances}). 
In agreement with previous simulations with depth-invariant flows, we find that within the interior the turbulent regime is predominantly characterized by the passive horizontal advection of each of the flow quantities (vertical vorticity, vertical velocity and fluctuating temperature). 
Details of the evolution of the flow quantities are best viewed by following fluid elements in a Lagrangian framework. This captures the hierarchy of secondary forces which drive the horizontal material advection, as defined by $D_t$.
%characterized by a predominance of the time derivative and advection terms in each of the three fluctuating eqtations. 
We find that the following analysis is best pursued utilizing a dimensional approach, followed by a reconversion to dimensionless quantities.
Considering the vertical momentum and vorticity equations in dimensional form, we find
\begin{align}
D^*_{t^*} w^* \sim \partial_{Z^*} p^* &>  g \alpha \theta^* \sim   \nu\nabla^{*2}_\perp w^*, 
\label{eq:wdim}\\
D^*_{t^*} \zeta^* \sim 2\Omega \partial_{Z^*} w^* & >  \nu\nabla^{*2}_\perp \zeta^*. 
\label{eq:zetadim}
\end{align}
%(Note that from our data, it is unclear whether or not the terms on the left hand side of equations \ref{eq:wdim}, \ref{eq:zetadim} are asymptotically larger than those on the right hand side in the limit of large $\Rat$. Thus, we opt to use the symbol $>$ rather than $\gg$.) 
We therefore find an interior Coriolis-Inertial (CI) balance in which the geostrophic pressure gradient and vortex stretching forces are the dominant secondary forces in the hierarchy. Furthermore, the rms buoyancy and rms viscous forces are comparable in magnitude and are the smallest of all terms. We note that these balances are consistent with those found in DNS studies of rotating convection \citep{aG21} and rotating convection-driven dynamos \citep{mY22}, \textcolor{black}{and they}
 imply that interior forcing and viscous dissipation play a subdominant role in the dynamics. Consequently, the observed dynamics must  be driven externally  by buoyant (A)rchimedean forces arising within the thermal boundary layer. 
%Indeed, force balances within the temperature equations indicates that thermal fluctuations behave as a passive scalar and obeys the advection-diffusion balance
%\be
%D^*_{t^*} \theta^* \sim  \kappa \nabla^{*2}_\perp \theta^*.
%\ee
 
 To summarize, we observe what may be termed a \textsl{non-local} CIA balance where the action of the Archimedean buoyancy force occurring within the thermal boundary layers is spatially separated from the interior Coriolis and inertial forces. It can now be shown that this finding resolves the dichotomy of not finding a diffusion-free integral scale $\ell^*_I$ as  predicted by a local CIA balance.   Assuming velocities achieve rotational free-fall, as presently observed, and that pressure is the geostrophic streamfunction, i.e., 
 \be
 \ub^*\sim U_{r\!f\!f} =\frac{g\alpha\Delta T}{2\Omega}, \quad  p^*\sim 2\Omega U_{r\!f\!f}\ell^*_I, \quad  \zeta^*\sim U_{r\!f\!f}/ \ell^*_I,
 \ee
  it follows from the  interior CI balance given in (\ref{eq:wdim},\ref{eq:zetadim})   that 
  \be
  %\nabla^*_\perp\sim  \frac{1}{\ell^*_I}, \quad 
  \partial_{Z^*}\sim \frac{1}{h^*}, \quad
  D^*_{t^*} \sim 2\Omega  \frac{\ell^*_I}{h^*}.
  \ee 
  Here $h^*< H$ is the vertical correlation height of interior convective motions.
 %, hereafter referred to nondimensionally as $h=h^*/H$. 
  Viscous processes are estimated via $\nabla^{*2}_\perp \sim (1/ \ell^{*}_d)^{2}$ where 
 %$\ell^*_d = \ell_d \ell^*_\nu$ 
 $\ell^*_d$  denotes the (optimal) dissipation length scale.
 %, recall $ \ell^*_\nu = E^{1/3} H$. 
With these findings the sub-dominance and equivalence of buoyancy and viscous forcing in (\ref{eq:wdim},\ref{eq:zetadim}) imply respectively
\be
\label{eq:intbal}
h^* < \frac{2\Omega}{\nu} \ell^*_I \ell^{*2}_d,\qquad
\frac{\theta^*}{\Delta_T} \sim \frac{\nu}{2\Omega  \ell^{*2}_d}
\ee
We note that $\theta^* = E^{1/3}\Delta_T \theta$ in the QG regime and that the boundary forced interior must also induce non-dimensional temperature fluctuations $\theta \sim  \lb\Rat/Pr\rb^{1/2}$  capable of transporting the dissipation-free heat flux in a manner consistent with the exact relation for the thermal dissipation rate \eqref{eq;powerlaw}, (see Figure~\ref{F:temp}\subref{temp_da}).
Equation \eqref{eq:intbal} can be reformulated non-dimensionally as
\be
   h  < \ell_I \ell^2_d, \quad \ell_d = \theta^{-1/2}
    \sim \lb \frac{\Rat}{Pr}\rb^{-1/4}
 \ee
with non-dimensional length scales $h= h^*/H$, $\ell_I=\ell^*I/\ell^*_\nu$ and $\ell_d=\ell^*d/\ell^*_\nu$, where $I$ and $d$ are non-dimensional coefficients; recall $ \ell^*_\nu = E^{1/3} H$.
% \be
%   \ell_I = \frac{\ell^*_I}{\ell^*_\nu} \gg 1, \quad \frac{\theta^*}{\Delta_T} \ll \ell_I E^{1/3} .
% \ee
Most importantly, we note that the absence of a local  Archimedean force in the interior balance places no dissipation-free restrictions on the scaling of the injection scale which is consistent with the simulated observation $\ell_I\propto\Rat^{1/8}$. It now follows that $h  \leq O(\Rat^{-3/8})$ indicating a shortening of vertical correlation length with increasing $\Rat$ as evident in Figure~\ref{temp} and \ref{vortz}.
 The interior force balances for the fluctuating temperature obeys 
 \be
 D^{\theta*}_{t^*} \theta^* \sim w^*\partial_{Z^*} \overline{\Theta}^* \sim \kappa \nabla^{*2}_\perp \theta^*.
\label{eq:tdim}
 \ee 
 such that unlike the momentum field, material advection (or stirring) of the temperature field is balanced by dissipation (mixing).
 \textcolor{black}{This can be observed by comparing the spatial morphologies of the midplane snapshots for temperature and vorticity in which the former quantity exhibits broader structures due to enhanced diffusion. }

The results presented in this study highlight the non-trivial behavior of rapidly rotating convection and provide a foundation for comparison with DNS and experiment in both the plane-layer and spherical geometries. By removal of the LSV, our study was able to investigate the origin of the inverse cascade and our results demonstrate how rotating convection departs from the theory of isotropic, homogenous turbulence. Most significantly, we have shown that diffusion free force-balance arguments, even in the regime of large $Re$ provide an incomplete picture of the convective dynamics, and are insufficient when attempting to identify the pertinent length scales in the flow. Instead, our investigations reveal that the interior, despite controlling the global momentum and heat transport, is forced externally by the boundary layers. 
Therefore, we \textcolor{black}{conclude} that QG dynamics at large Rayleigh number should not be considered diffusion free.

Several open questions are apparent from this investigation, \textcolor{black}{including:} Does the\\
geostrophic regime accessed in this study accurately represent the flow regime for $\Rat \to \infty$? If not, does a new, higher $\Rat$ regime exist in which the impact of molecular dissipation is diminished?
To what extent does domain geometry impact the ultimate scaling theory?
Investigating these questions remains challenging to laboratory studies and DNS given present difficulties in investigating broad ranges of the extreme geostrophic parameter regime.

\section*{Declaration of Interests}
The authors report no conflict of interest.

\section*{Acknowledgements}

The authors gratefully acknowledge funding from the National Science Foundation through grants EAR-1945270 (TGO, MAC), SPG-1743852 (TGO, MAC), and DMS-2009319 (KJ). The simulations were conducted on the Summit and Stampede2 supercomputers. Summit is a joint effort of the University of Colorado Boulder and Colorado State University, supported by NSF awards ACI-1532235 and ACI-1532236. Stampede2 was made available through Extreme Science and Engineering Discovery Environment (XSEDE) allocation PHY180013. %Visualizations were produced with VAPOR \citep{sL19}.
%Flow visualization was performed with VAPOR \citep{sL19}.  

%\appendix
%\section{}\label{appA}
\bibliographystyle{jfm}
%\bibliography{References,journal_abbreviations}%jfm-bib,References}
\bibliography{no_lsv.bbl}

\end{document}